\documentclass[sigconf]{acmart}

\usepackage{booktabs}
\usepackage{pifont}
\usepackage{amssymb}
\usepackage{balance}
\usepackage{graphicx}
\usepackage{amsmath,amssymb}
\usepackage{float}
\usepackage{dblfloatfix}
\usepackage{url}
\usepackage{multirow}
\usepackage{multicol}
\usepackage{color}
\usepackage{mathtools}
\usepackage{hyperref}
\usepackage{enumitem}
\usepackage[super]{nth}
\usepackage[caption=false]{subfig}
\usepackage[font=small,labelfont=bf]{caption}
\usepackage[export]{adjustbox}
\usepackage[linesnumbered,ruled]{algorithm2e}
\usepackage{soul}
\usepackage{rotating}
\hypersetup{
  colorlinks = true, 
  urlcolor   = blue, 
  linkcolor  = blue, 
  citecolor  = red   
}

\setlist[enumerate]{leftmargin=*,topsep=0pt,itemsep=0ex}
\let\emptyset\varnothing

\newcommand{\eg}{\textit{e.g., }}
\newcommand{\ie}{\textit{i.e., }}
\newcommand{\myitem}[1]{\vspace*{0.04in}\noindent\textbf{#1}}
\newtheorem{definition}{Definition}

\begin{document}
\title[O Peer, Where Art Thou?]{O Peer, Where Art Thou?\\Uncovering Remote Peering Interconnections at IXPs}

\author{George Nomikos}
\affiliation{
  \institution{FORTH, Greece}
}
\email{gnomikos@ics.forth.gr}

\author{Vasileios Kotronis}
\affiliation{
  \institution{FORTH, Greece}
}
\email{vkotronis@ics.forth.gr}

\author{Pavlos Sermpezis}
\affiliation{
  \institution{FORTH, Greece}
}
\email{sermpezis@ics.forth.gr}

\author{Petros Gigis}
\affiliation{
  \institution{FORTH / University of Crete, Greece}
}
\email{gkigkis@ics.forth.gr}

\author{Lefteris Manassakis}
\affiliation{
  \institution{FORTH, Greece}
}
\email{leftman@ics.forth.gr}

\author{Christoph Dietzel}
\affiliation{
  \institution{DE-CIX / TU Berlin, Germany}
}
\email{christoph@inet.tu-berlin.de}

\author{Stavros Konstantaras}
\affiliation{
  \institution{AMS-IX, Netherlands}
}
\email{stavros.konstantaras@ams-ix.net}

\author{Xenofontas Dimitropoulos}
\affiliation{
  \institution{FORTH / University of Crete, Greece}
}
\email{fontas@ics.forth.gr}

\author{Vasileios Giotsas}
\affiliation{
  \institution{Lancaster University, England}
}
\email{v.giotsas@lancaster.ac.uk}

\begin{abstract}
Internet eXchange Points (IXPs) are Internet hubs that mainly provide the switching infrastructure to interconnect networks and exchange traffic. While the initial goal of IXPs was to bring together networks residing in the same city or country, and thus keep \textit{local traffic local}, this model is gradually shifting.
Many networks connect to IXPs without having physical presence at their switching infrastructure. This practice, called \textit{Remote Peering}, is changing the Internet topology and economy, and has become the subject of a contentious debate within the network operators' community. However, despite the increasing attention it attracts, the understanding of the characteristics and impact of remote peering is limited. In this work, we introduce and validate a heuristic methodology for discovering remote peers at IXPs. We 
(i) identify critical remote peering inference challenges,
(ii) infer remote peers with high accuracy ($>$95\%) and coverage (93\%) per IXP, and 
(iii) characterize different aspects of the remote peering ecosystem by applying our methodology to 30 large IXPs.
We observe that remote peering is a significantly common practice in all the studied IXPs; for the largest IXPs, remote peers account for 40\% of their member base. We also show that today, IXP growth is mainly driven by remote peering, which contributes two times more than local peering.
\end{abstract}


\settopmatter{printacmref=false}
\setcopyright{none}
\renewcommand\footnotetextcopyrightpermission[1]{}
\pagestyle{plain}

\maketitle

\section{Introduction}
\label{sec:introduction}

Internet eXchange Points (IXPs) are crucial components of today's Internet ecosystem~\cite{ager2012anatomy, augustin2009ixps,chatzis2013benefits,chatzisimportance}, that provide infrastructure for the direct interconnection (\textit{peering}) of Autonomous Systems (ASes). Currently, there exist more than 700 IXPs around the world, with more than 11K member networks (i.e., \textit{peers}); these correspond to approximately 20\% of the total number of ASes~\cite{he_dataset,pch_dataset, peeringdb}. The largest IXPs host more than 800 networks each~\cite{amsixam_members,decixfr_members}, and handle aggregate traffic that peaks at or exceeds 6 Tbps~\cite{amsixam_stats,decixfr_stats}.

IXPs were originally created to locally interconnect ASes at layer-2 (L2), and \textit{keep local traffic local}~\cite{chatzis2013there}. Under this model, networks peer at IXPs to \textit{directly connect} with each other and avoid connections through third parties, and thus reduce costs, improve performance (\eg{lower latency}), and better control the exchanged traffic~\cite{ahmed2017peering,DrPeering-remote-peering}. However, the ever-increasing traffic flowing at the edge of the Internet, creates pressure for denser and more diverse peering that challenges the traditional IXP model. As a result, the IXP ecosystem is undergoing a fundamental shift in peering practices to respond to these requirements: networks may establish peering connections at IXPs from \textit{remote} locations, to broaden the set of networks they reach within one AS-hop~\cite{chiu2015we, roughan201110}, either over a (owned or rented) ``long cable'' or over \textit{resellers} that provide ports on the IXP and L2 access through their own network~\cite{ixreach_rp_service,retn_rp_service}. This practice contradicts the traditional view of IXPs as local hubs of direct peering and is commonly referred to as \textit{Remote Peering}~\cite{DrPeering-remote-peering} by IXP operators, where ``remote'' denotes a distant and/or indirect IXP connection: 

\begin{definition}
Remote Peering (RP) is when a network peers at an IXP without having physical presence in the IXP's infrastructure and/or through a reseller.
\label{def:remote-peering}
\end{definition} 

While RP has been actively advertised by IXPs, it has also fired up a heated debate within the operators' community ~\cite{euroix2016, DrPeering-remote-peering}. The \textit{proponents} of RP highlight the benefits in connectivity and cost reduction for the IXP members, whereas the \textit{opponents} emphasize on the risks and implications for network performance and resilience. Irrespective of which side in this debate one stands for, the reality is that RP is fundamentally changing the IXP peering landscape, with unclear effects on Internet economics and performance. Today we lack the tools and techniques to answer even simple questions, such as \textit{``Which peers of an IXP are remote and which are local?''}. The answer to this  question could significantly benefit Internet operations and drive routing policies and peering decisions (\eg{} eyeballs or content providers that seem local at an IXP may not be local).
Such knowledge is therefore important both for IXP operators to understand the characteristics of their member base, and IXP members to perform \eg{} traffic engineering (TE) based on peering policies. Moreover, it enables researchers to explore different facets of RP ecosystems.

In this paper, we propose a methodology to infer RP, and analyze its main characteristics. 
Our primary objective is to enable transparency, a property which is desired by all stakeholders, regardless of which side they pick in the RP debate.
We first provide the necessary background on this debate as well as related work in Section~\ref{sec:background}. After presenting our measurement datasets (Section~\ref{sec:datasets}), we make the following contributions:

\myitem{Identify inference challenges (Section~\ref{subsec:rtt-limitations}).} We first identify the difficulties of inferring RP, by collecting and analyzing a best-effort validation dataset of remote/local peers in 15 large IXPs. We show that inference based exclusively on latency measurements, as proposed by Castro \emph{et al.}~\cite{castro2014remote}, is not capable of accurately inferring RP at scale.

\myitem{Infer remote peers (Section~\ref{sec:inf-methodology}).} We design a novel methodology to infer whether a peer is remote or local to an IXP. Due to the involved complexity and challenges, we take into account multiple dimensions of peering, such as latency, colocation and IXP facility information, IXP port capacity and router connectivity, and combine them to achieve an accurate inference. Comparing our inferences against validation data shows that our approach achieves a 95\% accuracy and 93\% coverage, while the corresponding percentages of the state-of-the-art~\cite{castro2014remote} are 77\% and 84\%, respectively.

\myitem{Characterize remote peering (Section~\ref{sec:results}).} We apply our methodology to {30} large IXPs, and analyze characteristics of RP. 
While an extensive evaluation of RP characteristics and implications is outside the scope of this work, we consider use cases that exhibit the applicability of our inference approach.
We find that RP is prevalent today, with {28\%} of the peers being remote. Our results also show that today, IXP growth is mainly driven by remote peering, which contributes \textit{two times more} than local peering with respect to the number of new IXP members.

We further discuss relevant insights which arise from our study, including potential implications of RP (Section~\ref{sec:discussion}). Finally, we describe follow-up research directions, such as traffic analysis and a large-scale longitudinal study (Section~\ref{sec:future-work}).

\section{Background \& Related Work}
\label{sec:background}

\myitem{Peering at IXPs.} ASes connect and exchange traffic (\ie{\textit{peer}}) with each other via bi- or multi-lateral setups at IXPs, which operate L2 switching platforms. Typically, ASes become \textit{members} of an IXP by connecting to its infrastructure through their own router(s), colocated at the facility where the IXP has presence. This enables them to peer with other IXP members.

\myitem{Remote peering at IXPs.}
Remote peering does not require physical presence of networks' routing equipment in the IXP fabric~\cite{nipper2016}. The connection is performed through: (i) \textit{resellers}~\cite{souquet2016} of IXP ports that connect the remote peer's router(s) to the IXP switches, (ii) \textit{L2 connections}  (``long cables'') to the IXP facility (with ports bought by the peer itself), either with privately owned cables or by using a carrier, and (iii) \textit{IXP federations}~\cite{amsix-easyaccess,decix-globepeer}, \ie {IXPs belonging to the same organization} (like DE-CIX Frankfurt and DE-CIX New York), which are interconnected so that local peers of one IXP are remote to the other and vice versa\footnote{The involved IXPs still use their own route servers and BGP communities and serve their own member base.}.

\myitem{Wide-area IXPs and Remote Peering.} Some IXPs are geographically distributed entities, possessing switching infrastructure in multiple facilities in different metropolitan areas\footnote{We consider as metropolitan area a disk with diameter 100 km.}/countries.
We call such cases, where the IXP's L2 network spans large geographical areas, \textit{wide-area} IXPs. An example of a wide-area IXP is NL-IX~\cite{nlix}, spanning the European continent\footnote{While IXPs such as DE-CIX may have presence in multiple cities (e.g., Frankfurt, New York), they are not considered as wide-area IXPs, since they operate an independent/separate IXP at each city. In contrast, NL-IX is a sole IXP entity with a network distributed among multiple countries/cities. 
}. The members of a wide-area IXP are local peers, as long as they are directly patched to the switching infrastructure of at least one facility of the IXP (see Definition~\ref{def:remote-peering}); otherwise they are remote. Note that such IXP setups can heavily complicate remote peering inferences (see Section~\ref{subsec:challenge_wide_area_ixps}).

\myitem{The remote peering debate.}
The increasing attention that remote peering is drawing has also given rise to a recent debate within the networking community~\cite{euroix2016}, placing emphasis on the impact of remote peering on Internet routing and economics.

\myitem{Remote peering is good!}
There are several advantages and new possibilities for \textit{networks} peering remotely:
\begin{itemize}[leftmargin=*]
\item \textit{Monetary savings.} CAPEX is reduced since there is no need for additional routing equipment, or colocation and installation fees~\cite{DrPeering-remote-peering,Norton-remote-peering}. Remote peering can also be an option for offloading transit traffic~\cite{castro2014remote}.
\item \textit{Increased connectivity.} Networks can easily establish direct connections with more peers (\eg{content providers present at remote IXPs}), and have better control over traffic routed from / towards them.
\end{itemize}
For the \textit{IXP}, remote peering leads to:
\begin{itemize}[leftmargin=*]
\item \textit{More members/customers.} IXPs can attract members which are present in different cities or countries, and thus, increase their market share. IXPs with many members are more visible and appealing to potential customers.
\item \textit{Reseller ecosystem.} The IXP can benefit from reseller organizations, which handle new IXP memberships at scale, and therefore the setup and billing of new members is simplified.
\end{itemize}

\myitem{Remote peering is bad!} On the other hand, some network and IXP operators claim that remote peering is a disservice to the Internet~\cite{euroix2016,DrPeering-remote-peering}. IXPs have been originally created as peering hubs to keep ``\textit{local traffic local}''~\cite{chatzis2013there}. Changing this trend might lead to:

\begin{itemize}[leftmargin=*,topsep=0pt]
\item \textit{Degradation of performance.} Links over IXPs involving peers at distant locations from IXPs are expected to have larger latency (RTTs) than links between local peers. Hence, direct peering connections on IXPs might not necessarily lead to improved quality in communication. Additionally, resellers usually offer low capacity IXP ports (\eg{100Mbps}; see Section~\ref{subsec:port-capacity-feature}), which can cause congestion~\cite{fanou2017investigating}.

\item \textit{Loss of resilience.} While a network might have separate L3 connections with its peers on an IXP, in the case of remote peering some of these connections might share a common port (\eg{resellers sell fractions of the same physical IXP port to multiple remote peers}). A single outage on this port can thus affect (a) multiple connections, and (b) networks hundreds or thousands of kilometers away from the IXP. As a result, neither traffic nor outages ``stay local''. 
\end{itemize}

\myitem{Need for transparency.} While there is no consensus on whether remote peering is a good or bad practice, both its proponents and opponents acknowledge the necessity for understanding the characteristics of remote peering. Network operators want to know which peers are local or remote, where they are located, and the implications on the communication (\eg{latency, bandwidth, resilience}) among peers. This knowledge is critical since it can guide traffic engineering and peering policies.

\myitem{Related work.} Prior works on IXPs explore various aspects of the IXP eco\-system and show its impact on the Internet's hierarchical topology~\cite{ager2012anatomy,augustin2009ixps}, traffic exchange economics~\cite{chatzis2015quo,lodhi2014open}, 
and content delivery~\cite{bottger2017hypergiant,chatzis2013there, Stocker2016Content}. Others discuss multilateral peering over IXPs at scale~\cite{giotsas2013inferring} and show that interconnection strategies, such as RP, and extensive colocation practices~\cite{giotsas2015mapping}, create unexpected interdependencies among peering infrastructures~\cite{giotsas2017detecting}. Other work investigates the impact of RP~\cite{gupta2014peering} on the topology or the performance of continental peering ecosystems, such as Africa~\cite{fanou2017investigating}. Castro \emph{et al.}~\cite{castro2014remote} aimed to explore the traffic offloading capabilities of RP and provided a simple RTT-based approach for inferring RP. 

However, in our work we show that RTT alone~\cite{castro2014remote} is not sufficient to achieve accurate inference (see Section~\ref{subsec:rtt-limitations}). Instead, we combine RTT measurements with several other domain-specific design aspects of remote peering and achieve significantly larger accuracy and coverage levels, calculated using a substantial validation dataset. Our goal is to establish a general, thoroughly validated RP inference methodology and yield valuable insights on the global RP ecosystem.

\section{Datasets \& Measurements}
\label{sec:datasets}

\subsection{Active Measurement Sources}
\label{subsec:msm_sources}

We employ ping measurements to estimate the latency (RTT) between an IXP and its member ASes, and traceroute measurements to extract the IP-level paths traversing peering links.

\myitem{Pings.} We conduct ping measurements from a number of Vantage Points (VPs), namely \textit{Looking Glasses (LGs)} and \textit{RIPE Atlas probes (RA)}; the exact location of these VPs is known. Castro et al.~\cite{castro2014remote} used the PCH LGs~\cite{pch_dataset} that provided access to PCH border routers deployed in 22 IXPs. Unfortunately, PCH does not allow ping queries through their LGs anymore. Instead, using IXP websites, we compiled a list of $23$ publicly accessible LGs, that provide direct interfaces inside the IXP networks, \eg to an IXP route server. To automate the querying of these LGs we use the Periscope platform~\cite{giotsas2016periscope}.

We augment the set of the ping-enabled VPs through RA~\cite{RipeAtlas}, a well-established global Internet measurement platform with more than 25,000 probes. To identify RA probes colocated with IXP infrastructure, we search for probes with source IPs in the address space of an IXP's peering LAN, 
and for probes which resolve to an ASN assigned to an IXP NOC\footnote{
Note that probes connected to the IXP members themselves are not useful for our methodology, since these members can be also remote to the IXP, and thus may affect the RTT-based inference step biasing the ping measurements.}.
We discovered $66$ such RA probes. 

Merging the available LG and RA VPs provides good coverage in the RIPE ($29$ IXPs) and APNIC ($11$ IXPs) regions. Only $6$ IXPs are covered in the ARIN and LACNIC regions, and none under AFRINIC.

\myitem{Traceroutes.} We collect all the publicly available RA IPv4 traceroute measurements (\ie{built-in and user-defined}) ~\cite{RipeAtlas}. In total, we study $3.15$ billion traceroute paths towards $600K$ IPs, probed between Jan. 2017 and Mar. 2018.
We use the collected traceroute paths to extract IP-level IXP crossings (see Section~\ref{subsec:traixroute} and steps 3, 4 of Section~\ref{subsec:inference_algorithm}), as well as private connections between ASes over facilities (see step 5 of Section~\ref{subsec:inference_algorithm}).

\begin{table}[t]
\small
\centering
\caption{Overview of the IXP (IPv4) dataset and contribution of each data source.}
\vspace{-3mm}
\label{tab:ixp-dataset}
\resizebox{\columnwidth}{!}{
\begin{tabular}{l|c|cc|c|cc}
\hline
\multicolumn{1}{c|}{Source} & \multicolumn{3}{c|}{IXP Prefixes} & \multicolumn{3}{c}{IXP Interfaces} \\ \hline
                            & Total       & Unique  & Conflicts & Total        & Unique  & Conflicts  \\ \hline
\textbf{Websites}           & 42          & 4       &          & 12409            & 24    &           \\
\textbf{HE}                 & 429         & 51       & 1 (.010 \%)        & 29866            & 7659 & 80 (.27 \%)         \\
\textbf{PDB}                & 638         & 187       & 1 (.005 \%)        & 22146            & 1162       & 62 (.28 \%)          \\
\textbf{PCH}                & 467         & 129       & 1 (.007 \%)        & 5922            & 256       & 22 (.37 \%) \\ \hline
\textbf{Total}              & \textbf{731}  &         &           & \textbf{31690}   &         &            \\ \hline
\end{tabular}
}
\end{table}

\subsection{IXP Peering LANs and Ports}
\label{subsec:ixp_info}
 
Our methodology combines multiple sources of IXP-related information with the measurements of Section~\ref{subsec:msm_sources}.

\myitem{IXPs, members, and interfaces.} 
To identify traceroute hops that traverse IXPs, and feed our methodology with IXP-related information, we combine multiple sources to build an 
up-to-date list of \textit{IXPs}, their \textit{members}, and the \textit{associated IXP interfaces} (\ie{IP addresses belonging to IXP prefixes that are assigned to IXP member ASes}). We retrieve the related IXP information directly from IXP websites by parsing the provided Euro-IX~\cite{ixp_db} \texttt{json} and/or \texttt{csv} machine-readable formats, and the publicly available databases of Hurricane Electric (HE)~\cite{he_dataset}, PeeringDB (PDB)~\cite{peeringdb}, and Packet Clearing House (PCH) \cite{pch_dataset}.

To address cases of conflicting data, we consider IXP websites as the most reliable source of information since the data are directly provided by the IXP operators;
in fact, while websites may share peering policy information with e.g., PeeringDB, they maintain their own IXP-related information, such as membership lists. We then rank the other IXP sources based on their fraction of conflicting entries compared to the website data (Table~\ref{tab:ixp-dataset}). 
Consequently, we apply the following preference ordering to resolve conflicts: $IXP~websites>HE>PDB>PCH$. 

The final dataset includes $31,690$ IXP IP-to-AS mappings (\textit{IXP interfaces}) and $729$ IXP prefixes from $703$ IXPs (Table~\ref{tab:ixp-dataset}).
Interestingly enough, the IXP prefixes and interfaces that are unique in the websites are quite few (4 and 24 respectively), since the other databases are usually populated with up-to-date entries.
To the best of our knowledge, the collected dataset comprises the most complete list of IXPs, IXP prefixes, and IXP interfaces to-date.

\myitem{IXP port capacity.} 
We record the capacity of the peering ports allocated to each IXP member, using the \texttt{json}/\texttt{csv} datasets directly provided through the IXP websites, and the PDB records.
For each IXP, we also compile the available port capacity options through the pricing section of its website~\cite{ixpcost}. 
As we explain in Section~\ref{subsec:port-capacity-feature}, knowing the port capacities allows us to distinguish IXP peers that obtain virtual ports through port resellers from peers that obtain physical ports directly from the IXP.

\subsection{Detecting IXP Crossings in Traceroutes}
\label{subsec:traixroute}

We process traceroute measurements (Section~\ref{subsec:msm_sources}) and IXP information (Section~\ref{subsec:ixp_info}) with \texttt{traIXroute}~\cite{traixroutegit, nomikos2016traixroute} to identify paths that cross IXPs. We configure \texttt{traIXroute} to identify IXP crossings in a path, when (i) there exists a sub-path of three IPs (\ie{\textit{IP triplet}}) that contains an IXP IP in the middle of the triplet and this IXP IP belongs to the same AS as the 3$^{rd}$ IP, (ii) the AS of the 1$^{st}$ IP in the triplet is different,
and (iii) these two ASes are members of the IXP (whose prefix the IXP IP of the triplet belongs to).

\subsection{Colocation Facilities}
\label{subsec:fac_info}


\begin{figure}[t]
  \subfloat[Distribution of ASNs and IXP facilities][Distribution of ASNs and IXP\\facilities (source: PDB/Inflect).]{
  \includegraphics[clip,width=0.5\columnwidth,valign=t]{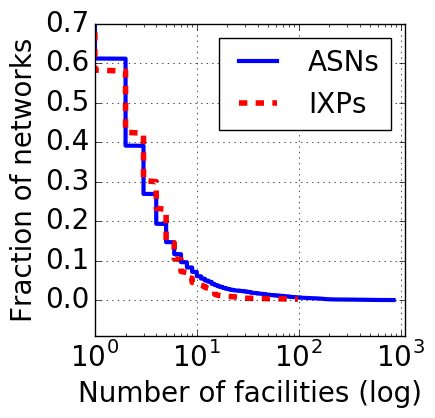}
  \label{fig:facilities_distribution}
  }
  \subfloat[ECDF of minimum RTTs for remote and local peers in the 
  control validation dataset. $18\%$ of the remote peers have RTT of less than 1ms while $40\%$ have RTT of less than 10ms.]
  {
  \includegraphics[width=0.5\columnwidth,valign=t]{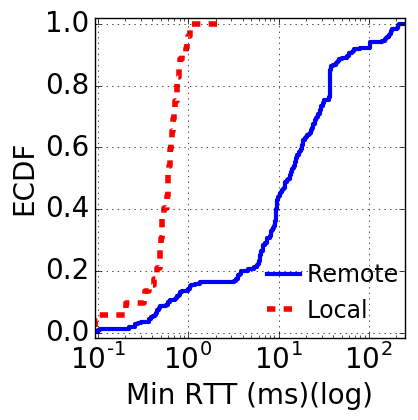}
  \label{fig:ground-truth-rtt}
  }
  \vspace{-3mm}
  \caption{Overview of facilities and VP-to-IXP interface RTT.}
  \label{fig:overall_facilities_distribution}
\end{figure}

To infer the remoteness or locality of peers, we also use the location of the facilities where IXPs and their members are present. We first collect the facility list from PDB and \textit{Inflect}~\cite{inflect}, a database for Internet infrastructure services
(whose data comes either directly from service providers or trusted third-party sources).
For each facility we keep the geographical coordinates 
provided by PDB, which are independently verified through \textit{Inflect} to filter-out spurious information~\cite{pdb_update_2017}.
Our dataset includes 656 IXPs which are associated with 1,078 facilities.
The Inflect dataset allows us to correct the geographical information for 308 of these facilities.
Moreover, we extract information related to which facility each AS (i.e., IXP member) is present.
As shown in Fig.~\ref{fig:facilities_distribution}, around 60\% of IXPs and ASes are present in a single facility, with only 5\% in more than 10 facilities. To alleviate possible incompleteness in PDB/Inflect data, we extend the colocation dataset by \emph{manually} extracting the facility list from the websites of the 50 IXPs with most AS members.
IXP websites provide additional facility data for 48\% of the IXPs, allowing us to compile 
an as complete as possible dataset for the most prominent IXPs. 

\myitem{PDB \textit{vs.} Websites.} We have encountered some discrepancies between PDB and IXP/facility websites. For example, the NL-IX website provides additional information on 17 ($\sim$15\%) of its data centers not present in PDB (incompleteness). 
On the other hand, for the CoreSite LA1 facility, PDB reports 108 ASes ($\sim$43\%) that are not listed in Coresite's list of locally deployed networks~\cite{coresite_carrier_list}, indicating possible inaccuracies in PDB. 
Even in the face of such artifacts, the combination of the heuristics we apply in Section~\ref{sec:inf-methodology} results to high accuracy/coverage.

\subsection{IXP Local/Remote Members for Validation}
\label{subsec:groundtruth_remote_members}

\begin{table}[t]
	{\small
    
	\caption{\small
    Validation data retrieved from IXP operators (top, 6 rows) and websites (bottom, 9 rows). IXPs with superscript '\emph{C}' ('\emph{T}') are part of the ``control'' (``test'') subset.
    }
    \vspace{-3mm}
	\label{table:validation-dataset}
		\begin{center}
        	\resizebox{\columnwidth}{!}{
			\begin{tabular}{clrrrrr}
				\toprule
				 & \textbf{IXP}      & \#\textbf{Facilities}      & \textbf{\#Total Peers}     & \#\textbf{Validated Peers} & \#\textbf{Local}  & \#\textbf{Remote} \\
				\midrule
				\multirow{6}{*}{\begin{turn}{90} \textbf{\textit{OPERATORS}} \end{turn}} 	 
				& AMS-IX$^T$         & 14  & 878    &     463  &   258  & 205 \\
				& DE-CIX FRA$^T$     & 28  & 795    &     323  &   103  & 220 \\
				& LINX LON$^T$       & 15  & 770    &     170   &    71  & 99 \\
     			& DE-CIX NYC$^C$     & 25  & 162    &     80   &    59  & 21\\
				& LINX MAN$^T$       & 3   & 99     &     37    &    17  & 20\\
				& LINX NoV$^T$       & 4   & 48     &     21    &     12  & 9\\
				\hline
				\multirow{9}{*}{\begin{turn}{90} \textbf{\textit{WEBSITES}} \end{turn}} 	
				& EPIX KAT$^C$       & 3   & 465     &     233   & 135 & 98  \\
				& EPIX WAR$^C$     & 6  & 308     &     170   &  93 & 77  \\
				& France-IX PAR$^T$  & 9  & 402     &     292   & 127 & 165  \\
				& Seattle IX$^T$     & 11  & 296    &     246    & 180 & 66  \\
		        & Any2 LA$^T$        & 2  & 299     &     212    & 147 & 65 \\
				& D. Realty ATL$^C$  & 3  & 142     &     85   & 42 & 43  \\
				& France IX MRS$^C$  & 2    & 77     &     31   & 19 & 12  \\
				& AMS-IX HK$^C$      & 2   & 46     &     24   & 14 & 10  \\
				& AMS-IX SF$^C$      & 4   & 36     &     23    & 16 & 7  \\
				\bottomrule
				& Total          & 131    & 4823    &     2410    &  1293 & 1117 \\
				\bottomrule
			\end{tabular}
            }
		\end{center}
	}
\end{table}

Inferring remote peering accurately, requires thorough investigation of the challenges related to interconnectivity between IXPs and their members, as well as information to validate the peering inference itself.
To this end, we contacted IXP operators and requested lists specifying which of their members are local and/or remote. We received
validation data\footnote{The validation dataset we use is a best-effort collection of relevant trusted data.} for 6 IXPs. However, the provided lists do not cover the entire list of the members of these IXPs. This is due to the fact that IXP operators usually know whether their members are connected through resellers, but not where they are located, or if they use a L2 carrier to access their colocation facilities. In essence, they do not/cannot know ``what goes on beyond that cable''~\cite{euroix2016}; a gap that is the primary motivation of this work.

We further augmented the
validation dataset by manually extracting lists of remote and local members from websites of IXPs that publish the port type of their members (physical or virtual through a reseller).
In total, we collected
validation data for 6 IXPs directly from their operators, and for 9 more IXPs from their websites. In addition, we enriched the total IXP list in the validation dataset with the facilities at which the IXPs are present based on data from Section~\ref{subsec:fac_info}.
All relevant statistics are shown in Table~\ref{table:validation-dataset}. 

We split the
validation dataset in two subsets, ``\textit{control}'' 
and ``\textit{test}'', 
depending on whether they include IXPs with publicly accessible colocated VPs from which ping measurements can be triggered. The reason for this discrimination is that we need to (i) re-evaluate existing inference approaches~\cite{castro2014remote} and investigate further challenges in order to fine-tune our approach, and (ii) properly validate the full methodology using active measurements.
Since only the \textit{test} subset contains IXPs with accessible ping-enabled VPs, we used the \textit{control} subset to evaluate latency-wise inference challenges (see Section \ref{subsec:rtt-limitations}), and the \textit{test} subset to ping local and/or remote target interfaces in order to compare our inference results with the reported ones (see Section~\ref{subsec:validation}).

\section{RTT-based Inference Challenges}
\label{subsec:rtt-limitations}

Here, we use  
the \emph{control} subset of our validation dataset to investigate the challenges and limitations of inferring RP based exclusively on latency measurements (Section~\ref{subsec:challenge_only_rtt}), placing emphasis on the fairly common case of wide-area IXPs (Section~\ref{subsec:challenge_wide_area_ixps}).

\subsection{RTT is not enough}
\label{subsec:challenge_only_rtt}

For each IXP in our control dataset, there is no publicly available VP to execute RTT measurements, but we obtained one-time access to results from pings executed within the IXP infrastructure targeting the peering interfaces of all the remote and local members of the IXP.
We apply the \textit{TTL match} and \textit{TTL switch} filters proposed in~\cite{castro2014remote} to discard replies with TTL values less than the expected maximum (64 and 255 hops) that may indicate ping replies outside of the IXP subnet. 
We repeat the measurements every 20 minutes for two days, and we calculate the minimum RTT per IXP interface. 
As shown in Fig.~\ref{fig:ground-truth-rtt}, RTT values above $2ms$ are a very strong indication of remote peers, with $99\%$ of the local peers having RTT values less than 1ms. This result is consistent with previous works that exhibited that a delay of 1ms corresponds roughly to a distance of 100 km~\cite{katz2006towards,trammell2018revisiting}, approximating the coverage (i.e., disk diameter) of a single metropolitan area. However, low RTT does not necessarily mean that a peer is local.
Surprisingly, \textbf{18\% of the remote peers in our control dataset are within 1ms from the IXP}, while $40\%$ are within $10ms$, which is the 
\textit{``remoteness threshold''} used in~\cite{castro2014remote}.

\subsection{Wide-area IXP challenges}
\label{subsec:challenge_wide_area_ixps}

\begin{figure}[t]
	\subfloat[\small Median RTTs between the  facilities of NET-IX][\small Median RTTs between \\ the  facilities of the \\ wide-area IXP NET-IX.]{
		\includegraphics[clip,width=0.51\columnwidth,valign=t]{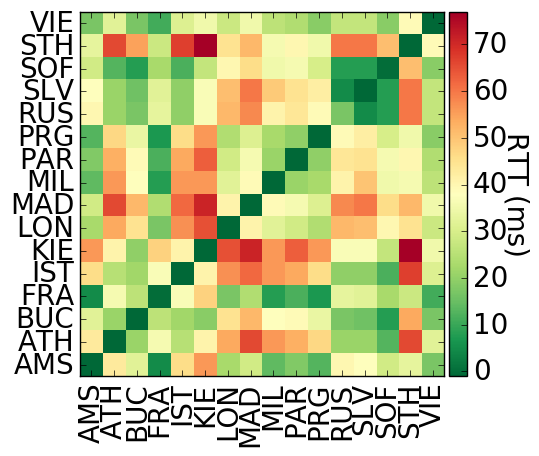}
		\label{fig:netix-rtts}
	}
	\subfloat[\small Maximum distance between IXP facilities, compared to the number of IXP members (source: PDB).][\small Max. distance between IXP\\facilities, compared to the\\number of IXP members\\(source: PDB).]{
		\includegraphics[width=0.5\columnwidth,valign=t]{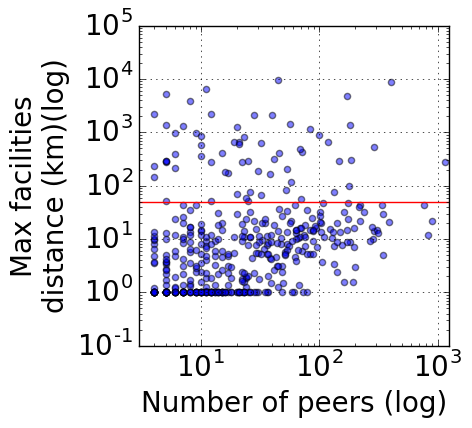}
        \label{fig:ixp-facilities-distance}
	}
    \vspace{-3mm}
	\caption{Features of wide-area IXPs.
}
\vspace{-1.2em}
\end{figure}
Conservative latency thresholds do not ensure the elimination of peers which are falsely identified as remote for \textit{wide-area IXPs}.
In fact, IXP members which are present in any of the facilities of such IXPs are local to the IXP but can be remote to the measurement VP, even if the VP is also hosted in one of the IXP's facilities.
An indicative example is NET-IX, which has distributed its switching fabric in facilities across 18 different countries~\cite{NetIXMap}.
To understand the RTT characteristics among the different facilities of such a geographically distributed IXP, we obtained pairwise delay measurements between 16 of NET-IX's international sites.
NET-IX measures the delay between its different facilities based on the Y.1731 Performance Monitoring standard~\cite{itu1731}, by sending precisely timestamped test packets across its MetroNID network demarcation points. 
The results are shown in Fig.~\ref{fig:netix-rtts}.
For 87\% of the facility pairs the median RTT is above $10ms$. 
Note that we also observe facilities in different countries with less than 10ms delay between them; for instance, Frankfurt (FRA) and Prague (PRA) have a $7ms$ delay. \textbf{Therefore, a remoteness RTT-threshold is not meaningful for wide-area IXPs.}

Next, we quantify the popularity of the model of the wide-area IXPs.
We use our colocation dataset compiled in Section~\ref{subsec:fac_info}, and we classify an IXP as wide-area if its switching fabric is deployed among multiple facilities, and at least two of them are in different metropolitan areas.
Since there can be different naming conventions used for the same city/metro area, we calculate the geodesic distance between each pair of IXP facilities, by applying Karney's method~\cite{karney2013algorithms} on their geographical coordinates.
We consider facilities more than $50km$ apart as located in different metropolitan areas.
For April 2018, we found that $64$ of the $446$ ($14.4\%$) IXPs in PDB with at least two IXP members are wide-area, including $10$ of the $50$ ($20\%$) largest IXPs in terms of the size of their IXP member list (Fig.~\ref{fig:ixp-facilities-distance}). \textbf{Therefore, wide-area IXPs are fairly common and not just some exceptional cases.}
Note that the infrastructure of some IXPs can be thousands of \textit{kms} apart. For instance, NL-IX has facilities in London and Bucharest that are over 1,300km away from each other. 

\myitem{}The results of this section highlight that although RTT measurements have the potential to provide useful insights w.r.t.~the peering approach employed by an IXP member, alone they are not adequate to accurately infer remote peers.
A $10ms$-threshold is very conservative in the case of IXPs concentrated in a single metropolitan area, while it yields a large number of false positives in the case of wide-area IXPs. 

\section{Inference Methodology} 
\label{sec:inf-methodology}

To address the limitations of remote peering inference based exclusively on latency measurements, we introduce a ``first-principles'' \cite{li2004first} approach. We rely on domain-specific knowledge to identify technological (beyond latency) and economic aspects of peering connectivity (Section~\ref{subsec:eng-inference}), and build upon these aspects to design a methodology for inferring remote and local peers (Section~\ref{subsec:inference_algorithm}). We validate the proposed methodology in Section~\ref{subsec:validation}.

\begin{figure*}[!ht]
\centering
	\subfloat[Multiple local IXP peerings.]{
		\includegraphics[clip,width=0.65\columnwidth,valign=t]{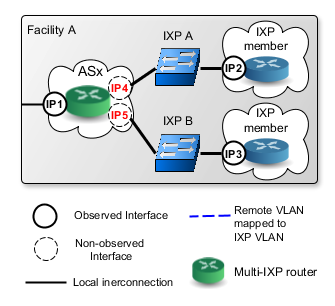}
		\label{fig:multirole-local}
	}
	\subfloat[Multiple remote IXP peerings.]{
		\includegraphics[clip,width=0.64\columnwidth,valign=t]{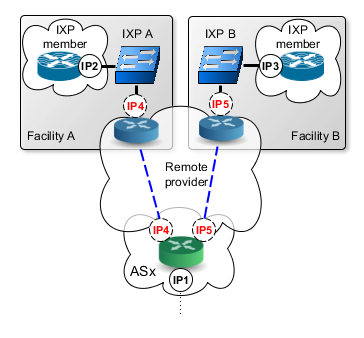}
		\label{fig:multirole-remote}
	}
	\subfloat[Local and remote IXP peerings.]{
		\includegraphics[width=0.55\columnwidth,valign=t]{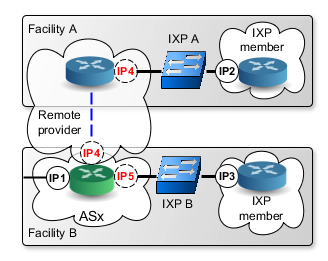}
        \vphantom{\includegraphics[width=0.17\textwidth,valign=t]{example-image-10x16}}
		\label{fig:multirole-hybrid}
	}
    \vspace{-3mm}
	\caption{Different scenarios of \textit{multi-IXP} routers, for which we may observe different traceroute paths where $IP_1$ precedes both IXP interfaces $IP_2$ and $IP_3$, indicating the presence of a multi-IXP router in $AS_x$.}
	\label{fig:multiixp-router}
    \vspace{-1.2em}
\end{figure*}
\begin{figure}[!t]
	\begin{center}
		\includegraphics[clip,width=7.5cm]{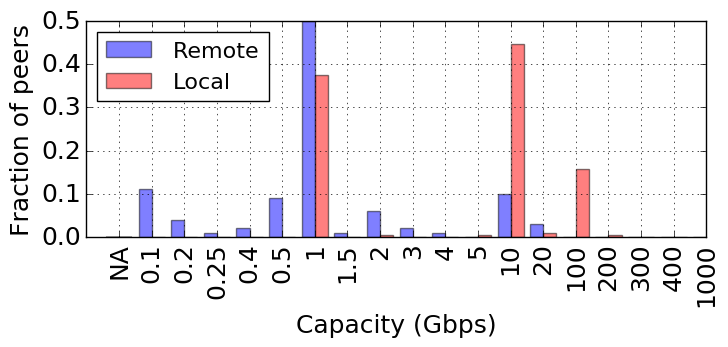}
		\vspace{-1em}
	\end{center}
	\caption{Capacity of IXP ports for remote and local peers in our
    control validation dataset. Fast Ethernet (FE) carries traffic at the rate of 100 Mbit/s and Gigabit Ethernet (GE) at 1 Gbit/s.}
	\label{fig:portspeeds-gt}
    \vspace{-1.2em}
\end{figure}

\subsection{Design Aspects}
\label{subsec:eng-inference}

\subsubsection{Port Capacity}
\label{subsec:port-capacity-feature}
IXPs offer to ASes connectivity to switch ports, whose capacity is typically between 1GE and 100GE~\cite{amsixam_pricing}. To make remote peering an attractive service, resellers split their physical ports to multiple virtual ports (e.g., via sub-interfaces/VLANs) of lower capacity (rate-limiting), and offer them to remote peers at lower prices. \textit{Fractional port capacities can be purchased only through resellers today}\footnote{\label{foot:old-members}In rare cases, some old IXP members are connected to physical ports of capacity less than the minimum offered today. This can be also due to stale entries in PDB.}. Thus, this information can indicate a network that peers remotely, via a reseller, at an IXP. Figure~\ref{fig:portspeeds-gt} shows the port capacity for remote and local peers 
in our control validation dataset. No local peer has port capacities below 1GE (which is the minimum capacity for physical ports offered by the corresponding IXPs), while 27\% of remote peers access the IXP through ports of 1FE -- 5FE capacity; on the other hand, ports of 100+GE are allocated only to local peers.

\subsubsection{Presence at Colocation Facilities}

\begin{figure}[t]
	\begin{center}
\includegraphics[clip,width=7.5cm]{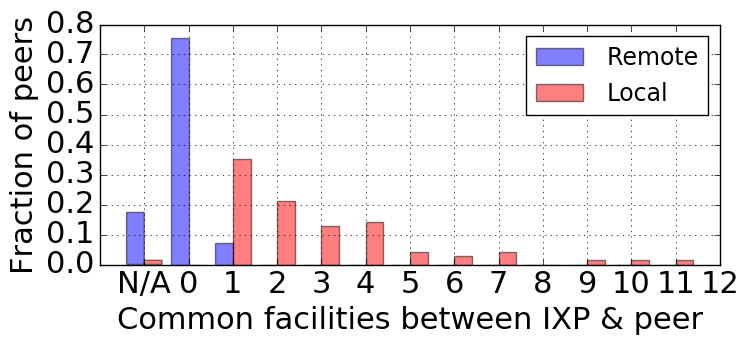}
		\vspace{-1em}
	\end{center}
	\caption{Number of IXP facilities where local and remote peers in our control 
    validation dataset are present.}
	\label{fig:facilities-overlap}
    \vspace{-1.2em}
\end{figure}

To establish a direct connection to an IXP, an AS needs to deploy routing equipment in at least one colocation facility where the IXP has deployed switching equipment. 
Therefore, \textit{it is not possible for an AS to be a local peer of an IXP if they are not colocated in a facility}. As Fig.~\ref{fig:facilities-overlap} shows, all local peers of an IXP in our control validation dataset are present in at least one IXP facility, while 95\% of the remote peers do not have any common facility with the IXP. 
Hence, assuming perfect knowledge of the facilities where IXP members are present, identifying RP would be a straightforward lookup process. However, the available colocation data for IXP members are incomplete and noisy. For example, in Fig.~\ref{fig:facilities-overlap}, there are no available data for $18\%$ of the remote peers, while $5\%$ of them appear to have presence in one IXP facility.

To further investigate the latter 5\% of RP cases, we contacted the IXP operators. Their feedback suggested that such cases are either an artifact of remote peers (not colocated with the IXP) adding the facility of their port reseller in their PDB record, or a consequence of the fact that peers (colocated with the IXP) prefer to connect through a port reseller in order to buy virtual ports of lower capacity at a discount price (see Section~\ref{subsec:port-capacity-feature}).

\subsubsection{Multi-IXP Routers}
\label{subsec:multi-ixp-feature}

An AS may connect to multiple IXPs through the same border router to reduce operational costs; we call such routers \textit{multi-IXP routers}. The IP interfaces of a multi-IXP router might appear in different traceroute paths to be interconnected with different IXPs.
We distinguish three cases where this is possible:
\begin{enumerate}
\item When multiple IXPs are present in the same facility, a colocated AS may connect directly to all of them using a single router (Figure~\ref{fig:multirole-local}).
\item Remote peers may connect through the same provider (port reseller) to multiple remote IXPs where this provider has presence (Figure~\ref{fig:multirole-remote}). 
\item An AS may connect with the same router to both local and remote IXPs, if it is e.g., colocated with one IXP and uses a reseller for another (Figure~\ref{fig:multirole-hybrid}).
\end{enumerate}

\subsubsection{Private Connectivity}
\label{subsec:private-connectivity}

Two networks colocated at the same IXP-hosting facility can interconnect with each other (\textit{private peering}) without using the IXP infrastructure, \eg{by directly connecting their routers}. This might be a more economical solution in case they exchange large volumes of traffic~\cite{norton2014}. 
Therefore, when an IXP member appears to be privately connected with several ASes which are colocated at the facility of the same IXP, this is a strong indication that this member is local to the IXP.

\subsection{Algorithm}
\label{subsec:inference_algorithm}

We next describe our methodology for inferring remote peering, by combining RTT measurements with the four peering aspects discussed in Section~\ref{subsec:eng-inference}.
While the steps of the methodology can be validated independently (see Section~\ref{subsec:validation}), the order in which they are applied matters and was selected as follows. Step 1 (Port Capacities) is first since it reliably infers RP, albeit with small coverage. Step 2 (RTT measurement) generates data used for step 3. Step 3 (RTT+colocation) is required as input by Step 4 (multi-IXP routers) and 5 (private connectivity). Step 4 comes before step 5 due to its higher accuracy; step 5 is the last resort for missing inferences.
Note also that while an individual step may miss some cases for different reasons (e.g., incomplete colocation data or RTT outliers in Step 2), these cases can be captured by a following step.

\paragraph{Step 1: Finding reseller customers via port capacities}
IXP members that reach the IXP through a reseller are identified as remote peers (see Definition~\ref{def:remote-peering}). As discussed in Section~\ref{subsec:port-capacity-feature}, members can be connected to IXP ports of capacity lower than the minimum physical port capacity $C_{min}$ offered by the IXP, only if they reach the IXP through a reseller\textsuperscript{\ref{foot:old-members}}. Hence, as a first step, for each IXP member $AS_{x}$ we compare the port capacity $C_{x}$, reported either in the IXP website or the Inflect and PDB databases, to the $C_{min}$ value reported in the pricing section of the IXP's website. If $C_{x}<C_{min}$, we infer that \textit{$AS_{x}$ is a remote peer} using a virtual port obtained through a reseller.

\begin{figure}[t]
	\begin{center}
		\includegraphics[clip,width=0.95\columnwidth]{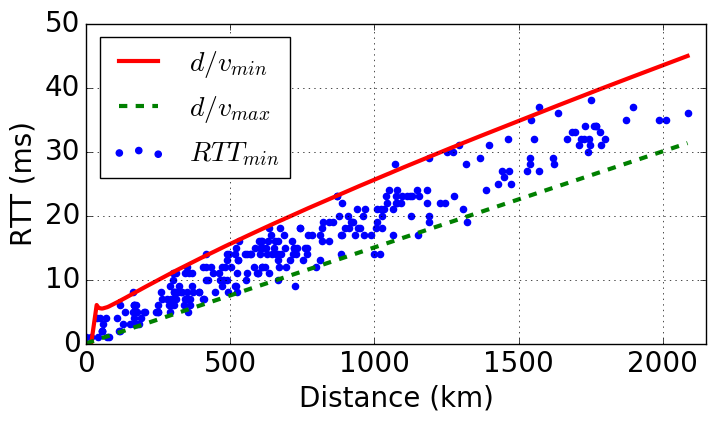}
		\vspace{-1em}
	\end{center}
	\caption{Inter-facility RTT as a function of distance, based on Y.1731 Perf. Monitoring measurements from NL-IX and NET-IX.}
	\label{fig:rtt-distance}
    \vspace{-1.2em}
\end{figure}

\paragraph{Step 2: Ping RTT Measurements}

From every VP in an IXP (see Section~\ref{subsec:msm_sources}), we execute ping measurements to every IXP IP interface of the IXP's members (see Section~\ref{subsec:ixp_info}). To reduce the sensitivity of the results to network conditions, we repeat the measurements every two hours for two days, which results in 24 measurements in total for each \{VP, IP interface\} pair. 
Similarly to Section~\ref{subsec:rtt-limitations}, we apply the \textit{TTL match} and \textit{TTL switch} filters to discard measurements without consistent TTL values. Finally, for each responsive IP interface we store the minimum RTT value, $RTT_{min}$,
to counter transient latency inflation artifacts~\cite{holterbach2015quantifying}.

\paragraph{Step 3: Colocation-informed RTT interpretation}

To infer local and remote peers, we analyze the collected $RTT_{min}$ values. Besides the colocation information of the IXPs and its members (see Section~\ref{subsec:fac_info}), the exact locations of the VPs are also known in all ping measurements. From the value of the $RTT_{min}$ we calculate a geographical area (circle or ring) around the VP location where the IP interface (and thus the router) of the IXP member can be located. The presence (or not) of a facility of the IXP in this area, denotes a local (or remote) peering, respectively.

\begin{figure}[t]
	\begin{center}
		\includegraphics[clip,width=0.95\columnwidth]{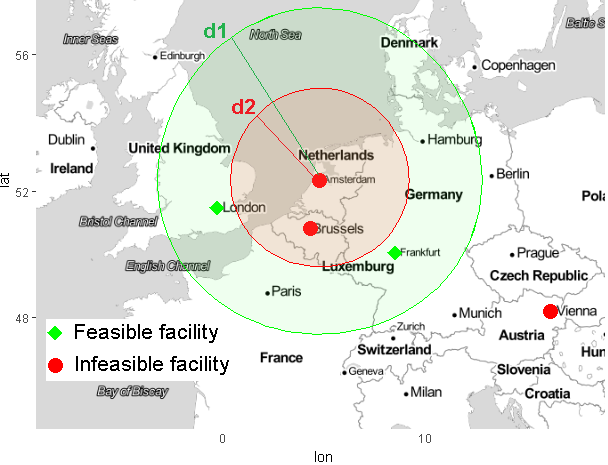}
		\vspace{-1em}
	\end{center}
	\caption{Example of combining RTT measurements with IXP colocation data to infer local peers at geographically distributed IXPs.}
	\label{fig:rtt-fac-example}
    \vspace{-1.2em}
\end{figure}

More precisely, we first calculate the distance between the involved VPs and each of the IXP's facilities, as described in Section~\ref{subsec:rtt-limitations}. Then, from the observed $RTT_{min}$, we calculate the potential distance between the VP and the ping target (IP interface at a member's router). Katz-Bassett \emph{et.~al}~\cite{katz2006towards} found that the end-to-end probe packet speed is at most $v_{max} = {\frac{4}{9}} \times c$, where $c$ is the speed of light. 
As shown in Fig.~\ref{fig:rtt-distance} (green/dashed curve), our dataset of facility-to-facility delays based on Y.1731 measurements obtained from NL-IX and NET-IX confirms this.
Through data fitting, we also find an approximate lower bound (red/continuous curve in Fig.~\ref{fig:rtt-distance}) for the speed $v_{min}(d) =10^7\cdot(ln(d)- 3)$, where $d$ is the distance.
Based on these bounds\footnote{ 
Out-of-bounds outliers do not impact the high accuracy of this step (see Table~\ref{tab:validation}).}, we estimate that the ping target is within a distance range $D_{feasible} = [d_{min}, d_{max}]$ (green area in Fig.~\ref{fig:rtt-fac-example}) from the VP, where $d_{min} = v_{min}\times{RTT_{min}}$ and $d_{max} = v_{max}\times{RTT_{min}}$. We call the facility that is located in $D_{feasible}$, a \textit{feasible facility}.

Based on the estimated area defined by $D_{feasible}$ (see \eg{Fig.~\ref{fig:rtt-fac-example}}), and the distances between the IXP facilities and the VP, we infer that the IXP member that owns the queried IP interface (ping target) is local or remote to the IXP, as follows:

\begin{enumerate}
\item \textbf{Remote peer}: if (i) the IXP has no available feasible facility, or (ii) the IXP has at least one feasible facility, but the peer is present in another feasible facility where the IXP is \textit{not} present.

\item \textbf{Local peer}: if the IXP has at least one feasible facility, and the IXP member is also colocated in one of the feasible IXP facilities.

\item \textbf{No inference}: if the IXP has at least one feasible facility, but the IXP member is \textit{not} present at any feasible facility. 
\end{enumerate}
In the latter case, it is likely that our colocation dataset is incomplete w.r.t.~the given peer. In this case we do not make an inference yet, but instead we leverage \textit{multi-IXP router} and \textit{private connectivity} information (see Section~\ref{subsec:eng-inference}) as described in the following steps.

Combining RTT values with colocation information allows us to alleviate false positives caused by wide-area IXPs.
Figure~\ref{fig:rtt-fac-example} shows an example of such a case, based on the topology of the NL-IX IXP.
The IXP has distributed its peering fabric across multiple cities, including Amsterdam, Brussels, London, Frankfurt and Vienna. Our measurement VP is in an IXP facility in Amsterdam, from which we ping the IXP peering interfaces.
Assume that for an interface $IP_x$ we measure an $RTT_{min}$ of $4ms$.
Without taking into consideration the geographical footprint of the IXP's infrastructure we would infer the corresponding peer as remote assuming a ``reasonable'' (see Fig.~\ref{fig:ground-truth-rtt}) 2$ms$-threshold. Instead, we find that the IXP has two feasible facilities (London and Frankfurt) in the ring between $d_1 = 532km$ and $d_2 = 299km$ from the VP, as defined by our $v_{max}$ (green area) and $v_{min}$ (red area) bounds respectively, allowing us to infer as local the IXP members colocated at these facilities.

Similarly, we can avoid false negatives due to remote peers that are in close proximity to the IXP. 
For instance, for a peer located in Rotterdam  connected remotely to the IXP's facility in Amsterdam ($57km$ distance) we will typically measure $RTT_{min} < 2ms$. By using the peer's collocation data we can correctly determine that, despite the low RTT, the peer is not local.

\paragraph{Step 4: Multi-IXP router inference}

The previous steps may not be able to infer the peering type due to missing facility data or missing RTT values from unresponsive IXP interfaces. In such cases, we proceed to use the multi-IXP router feature (see Section~\ref{subsec:multi-ixp-feature}), for inferring remoteness (or locality).

To identify multi-IXP routers we first collect traceroute paths from public RIPE Atlas measurements in the same period as our ping campaign (two days). We then extract the IP-level IXP crossings, as explained in Section~\ref{subsec:traixroute}, and we collect all sequences of hops $\{IP_{x}, IP_{x+1}^{IXP}\}$, where the interface $IP_{x+1}^{IXP}$ belongs to the address space of an IXP, and the interface $IP_{x}$ belongs to an AS that is a member of this IXP. For each AS that appears to peer at more than one IXP in different IXP crossings, we perform alias resolution on all its IP interfaces using MIDAR~\cite{keys2013internet} to map these interfaces to routers\footnote{
There are two available datasets based on MIDAR:
(i) one based on aliases resolved with MIDAR and \texttt{iffinder}~\cite{iffinder}, yielding the highest confidence aliases with very low false positives, and (ii) one also including aliases resolved with kapar~\cite{kapar}, which significantly increases coverage at the cost of accuracy. We selected the first dataset to \textit{favor accuracy over completeness}.}.
For interfaces on the same router, we find the set of IXPs that appear as next hops in traceroute paths. If a router appears to have connections to more than one IXPs, we characterize it as a multi-IXP router.

For example, assume two sequences of IP hops, $\{IP_{a}$, $IP_{IXP1}\}$ and $\{IP_{b}, IP_{IXP2}\}$, where both $IP_{a}$ and $IP_{b}$ are owned by the same AS and are mapped to the same router $R$, and $IP_{IXP1}$ and $IP_{IXP2}$ belong to the peering LANs of $IXP1$ and $IXP2$, respectively. In this case, $R$ has layer-3 connectivity with both IXPs, and therefore we characterize $R$ as a multi-IXP router. 

We then classify the multi-IXP routers in one of the categories described in Fig.~\ref{fig:multiixp-router}, and infer each one based on geolocation data from Section~\ref{subsec:fac_info} as follows:
\begin{enumerate}
\item \textbf{Local multi-IXP router}: A multi-IXP router is local to all involved IXPs (Fig.~\ref{fig:multirole-local}), if (i) the involved AS has been inferred as local peer --from previous steps-- in at least one of the IXPs, and (ii) the involved IXPs have at least one common facility. Then \textit{the AS is inferred as a local peer to all involved IXPs}.
\item \textbf{Remote multi-IXP router}: A multi-IXP router is remote to all involved IXPs (Fig.~\ref{fig:multirole-remote}), if (i) the involved AS has been inferred as remote peer --from previous steps-- in at least one of the IXPs (\eg{$IXP_R$}), 
and (ii) at least one of the following holds:
\begin{enumerate}
\item all the involved IXPs have at least one common facility. 
\item the maximum distance between the facilities of any involved IXP and $IXP_R$, is smaller than the minimum possible distance $d_{min}$ between all the facilities of the involved AS and all the facilities where $IXP_R$ is present.
\end{enumerate}
Then \textit{the AS is inferred as a remote peer to all involved IXPs}.

\item \textbf{Hybrid multi-IXP router}: A multi-IXP router is local to a subset of the involved IXPs (Fig.~\ref{fig:multirole-hybrid}) and remote to another IXP subset, if (i) the involved AS has been inferred as local peer --from previous steps-- in at least one of the IXPs (\eg{$IXP_L$}) of the local subset, and (ii) at least one of the following conditions is true for the remote subset:
\begin{enumerate}
\item $IXP_L$ does not have any common facility with the other involved IXPs.
\item the minimum distance between the facilities of $IXP_L$ and any other involved IXP, is larger than the maximum possible distance $d_{max}$ between all the --common-- facilities where both the involved AS and $IXP_L$ are present.
\end{enumerate}
Then \textit{the AS is inferred as a local peer to $IXP_L$ and remote peer to all other involved IXPs in the remote subset}.
\end{enumerate}

To understand the intuition behind conditions $2(b)$ and $3(b)$, assume that $R_x \in AS_x$ is a multi-IXP router peering with two IXPs, $IXP_{ams}$ in Amsterdam, and $IXP_{lon}$ in London.
The minimum distance between the facilities of the two IXPs is 300km, while the maximum distance is 360km. If from the first two steps we inferred that $AS_x$ is remote to $IXP_{ams}$, with $d_{min} = 500km$, then $R_x$ cannot be local to any facility of $IXP_{lon}$ (condition 2(b) holds). Similarly, if we inferred that $AS_x$ is local to $IXP_{ams}$ with $d_{max} = 50km$, then $R_x$ cannot be local to any facility of $IXP_{lon}$ (condition 3(b)  holds). 

\paragraph{Step 5: Localization of private connectivity}

If Steps 1-4 fail to infer whether a peer is local or remote, we use the private connectivity of an IXP member and apply a ``voting'' scheme similar to the Constrained Facility Search (CFS) approach~\cite{giotsas2015mapping}.

Let $\mathcal{F}_{IXP}$ be the set of feasible facilities for the IXP, $AS_{x}$ an IXP member identified based on the dataset of Section~\ref{subsec:ixp_info}, and $\mathcal{I}_{IXP}$ the set of all IP interfaces of the multi-IXP routers identified in Step 4.
\begin{enumerate}
\item We parse all the collected traceroute paths, perform IP-to-AS mapping~\cite{caidapfxtoas} and extract all the AS sequences over \textit{private interconnections} (not over an IXP), \ie{from a sequence \{$IP_{i}$,$IP_{j}$\}, where  $IP_{i}$ belongs to $AS_{i}$ and $IP_{j}$ to $AS_{j}$ ($ \neq AS_{i}$), we extract the sequence \{$AS_{i}$,$AS_{j}$\}}. 
Let $I_{priv}$ be the set of all interfaces involved in such private AS-level interconnections.
\item We run alias resolution on the interfaces in $I_{IXP} \cup I_{priv}$, that belong to IXP members for which we have not made an inference yet.
For each router $R_{x}$ (belonging to an $AS_{x}$) with at least one interface $i \in I_{IXP}$, we compile the set $\mathcal{N}_{x}$ of the (private) AS neighbors of $AS_{x}$.
\item Based on our AS-to-facility mapping from Section~\ref{subsec:fac_info},
we find the most common facilities $\mathcal{F}_{common}$ among the majority of the ASes in $\mathcal{N}_{x}$.
\end{enumerate}

If $|\mathcal{F}_{IXP} \cap \mathcal{F}_{common}|=1 $, \ie{only one facility of the IXP belongs to both sets}, then \textit{we infer $AS_{x}$ as a local peer to the $IXP$}. Otherwise \textit{we infer the peer as remote to the IXP}. 
The intuition behind this heuristic is that private interconnections are typically established within the same facility, as explained in~\ref{subsec:private-connectivity}. Nonetheless, we do not require all the private AS neighbors to be present in $\mathcal{F}_{common}$ because tethered private interconnections across facilities --although less common-- are still possible~\cite{giotsas2015mapping}.

It should be noted that our aim is not to pinpoint the exact AS boundaries, nor to derive the AS-level topology from IP hops, both of which have been shown to be non-trivial processes~\cite{luckie2016bdrmap,marder2016map}. Instead, we aim to infer a router's colocation facility based on its adjacent ASes. For example, a reply from a third-party interface may result in a spurious AS-level link; however, the interface (no matter to which AS it is mapped) belongs to the same router, and thus the facility inference is not affected.

\begin{figure}[t]
	\begin{center}
		\includegraphics[clip,width=0.9\columnwidth]{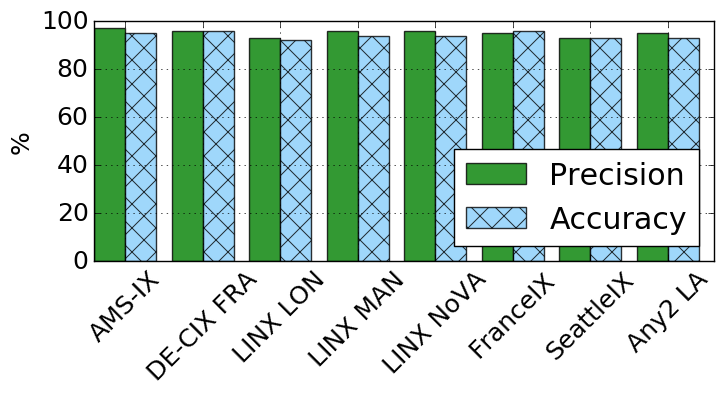}
		\vspace{-1em}
	\end{center}
    \vspace{-3mm}
	\caption{Validation results per IXP in our test validation dataset.}
	\label{fig:validation-per-ixp}
    \vspace{-1.2em}
\end{figure}

\subsection{Validation}
\label{subsec:validation}

\begin{table}[!b]
\centering
\small
\caption{
Validation Sets and Metrics for RP Inference.}
\vspace{-3mm}
\label{tab:val-def}
\begin{tabular}{c|c|l}
&Name&Definition\\
\hline
\multirow{8}{*}{Sets}&$\mathcal{VD}_{R}$&Remote Peers in Validation Dataset\\
\cmidrule{2-3}
&$\mathcal{VD}_{L}$&Local Peers in Validation Dataset\\
\cmidrule{2-3}
&$\mathcal{VD}$&$\mathcal{VD} = \mathcal{VD}_{R} \cup \mathcal{VD}_{L}$\\
\cmidrule{2-3}
&$\mathcal{INF}_{R}$&Inferred Remote Peers\\
\cmidrule{2-3}
&$\mathcal{INF}_{L}$&Inferred Local Peers\\
\cmidrule{2-3}
&$\mathcal{INF}$&$\mathcal{INF} = \mathcal{INF}_{R} \cup \mathcal{INF}_{L}$\\
\hline
\hline
\multirow{8}{*}{Metrics}&COV&$\frac{|\mathcal{INF}\cap \mathcal{VD}|}{|\mathcal{VD}|}$ (Coverage)\\
\cmidrule{2-3}
&FPR&$\frac{|\mathcal{INF}_{R}\cap \mathcal{VD}_{L}|}{|\mathcal{INF}\cap\mathcal{VD}_{L}|}$ (False Positives rate)\\
\cmidrule{2-3}
&FNR&$\frac{|\mathcal{INF}_{L}\cap \mathcal{VD}_{R}|}{|\mathcal{INF}\cap\mathcal{VD}_{R}|}$ (False Negatives rate)\\
\cmidrule{2-3}
&PRE&$\frac{|\mathcal{INF}_{R}\cap \mathcal{VD}_{R}|}{|\mathcal{INF}_{R}|}$ (Precision)\\
\cmidrule{2-3}
&ACC&$\frac{|\mathcal{INF}_{R}\cap \mathcal{VD}_{R}|+|\mathcal{INF}_{L}\cap \mathcal{VD}_{L}|}{|\mathcal{INF}|}$ (Accuracy)\\
\hline
\end{tabular}
\end{table}
\begin{table}[!t]
\centering
\small
\caption{Validation of each step of the algorithm.}
\vspace{-3mm}
\label{tab:validation}
\resizebox{\columnwidth}{!}{
  \begin{tabular}{rl|lllll}
  \hline
  \multicolumn{1}{c}{\textbf{Methodology}} & \textbf{Feature} & \textbf{FPR} & \textbf{FNR} & \textbf{PRE} & \textbf{ACC} & \textbf{COV} \\
 \cline{2-7}
 
 \multicolumn{1}{c}{\textbf{Steps}} & $ RTT_{min}$ \cite{castro2014remote} & 17.5\%        & 25.7\%       & 85\%         & 77\%      & 84\%            \\
  \hline
  \hline
  Step 1: & Port Capacity  & -             & -            &    96\%           &   -   & 11\%          \\
  \hline
  Step 2+3: & $RTT_{min}$+Colo & 1.1\%         &    7\%         & 98.5\%       & 95.6\%       & 76\%            \\
  \hline
  Step 4: & Multi-IXP &  7\%          &       7\%       &      93\%        &     93\%       & 53\%            \\
  \hline
  Step 5: & Private Links &      10\%        &      16\%        &          90\%    &    86.5\%        & 49\%            \\
  \hline
  & \textbf{Combined} &   \textbf{4}\%          &   \textbf{7.2}\%           &       \textbf{95}\%       &      \textbf{94.5}\%        &              \textbf{93}\%     \\ \hline
  \end{tabular}
}
\end{table}

We validate each step of our methodology independently by comparing inference results (see Section~\ref{subsec:step-inf-wild}) against the
\textit{test} subset of the validation dataset (see Section~\ref{subsec:groundtruth_remote_members}). 
The validation metrics we use and the sets that we consider are defined in Table~\ref{tab:val-def}.
Note that concerning validation data it holds that $\mathcal{VD}_{R}\cap\mathcal{VD}_{L}=\emptyset$ (on the interface level), and in the metrics we do not take into account inferences for peers with no validation data (\ie{$\mathcal{INF}-\mathcal{VD}=\emptyset$}).
Table~\ref{tab:validation} shows the validation results for all IXPs in the test dataset, for each step separately, as well as the entire algorithm.

\myitem{State-of-the-art.} As a baseline, we 
first validate the remote inference when using only $RTT_{min}$
(\textit{step 2}), assuming a remoteness threshold of $10ms$~\cite{castro2014remote}, to quantify the improvement versus the state of the art~\cite{castro2014remote} achieved by our algorithm. $RTT_{min}$ yields a high $FPR$ due to mis-inferring local peers at wide-area IXPs as remote. We calculated that when excluding wide-area IXPs the $FPR$ of the $RTT_{min}$ approach drops to 2\%.
At the same time, the $FNR$ is also high since many of the remote peers have $RTT_{min}<10ms$.

\myitem{Proposed methodology}. When combining $RTT_{min}$ with colocation data from Section~\ref{subsec:fac_info} (\textit{step 3}) we improve significantly all validation metrics; only the coverage metric has a small decrease, due to the fact that both latency and facility data are required. The false-negative inferences of $RTT_{min}+Colo$ are either due to spurious colocation data, or reseller customers colocated at the IXP. The latter false negatives are alleviated by taking into account \textit{Port Capacity} data as described in Section~\ref{subsec:ixp_info} (\textit{step 1}). For port capacity we validate only the precision metric, since we use it to infer only remote peers. For the next two steps we utilize traceroute data from Section~\ref{subsec:msm_sources}. Specifically, the $Multi-IXP$ step (\textit{step 4}) also exhibits very high PRE and ACC, but can be used only for half of the interfaces. Finally, the \textit{Private Links} step (\textit{step 5}) has the lowest ACC and PRE compared to the other steps, but still outperforms vanilla RTT-based inference and is used only as a ``last-resort'' heuristic. 
When all the five steps are combined, they yield $\sim$95\% ACC and PRE, and cover 93\% of the tested IXP interfaces.
Fig.~\ref{fig:validation-per-ixp} shows the precision and accuracy metrics per IXP in our test validation dataset, ordered by the size of IXP. The results are consistent across all IXPs. For SeattleIX we obtain the lowest precision (92\%), due to incomplete colocation data. Our inferences for LINX LON have the lowest accuracy (91\%), because of a higher --than the other IXPs-- number of colocated members connected through remote providers using non-fractional ports. These inaccuracies may indicate potential errors in the port capacities dataset.

\section{Inferring RP in the Wild}
\label{sec:results}

\begin{figure*}[!ht]
\centering
\subfloat[Response rate of LGs and Atlas Probes.  ]{
		\includegraphics[clip,valign=t,width=4cm]{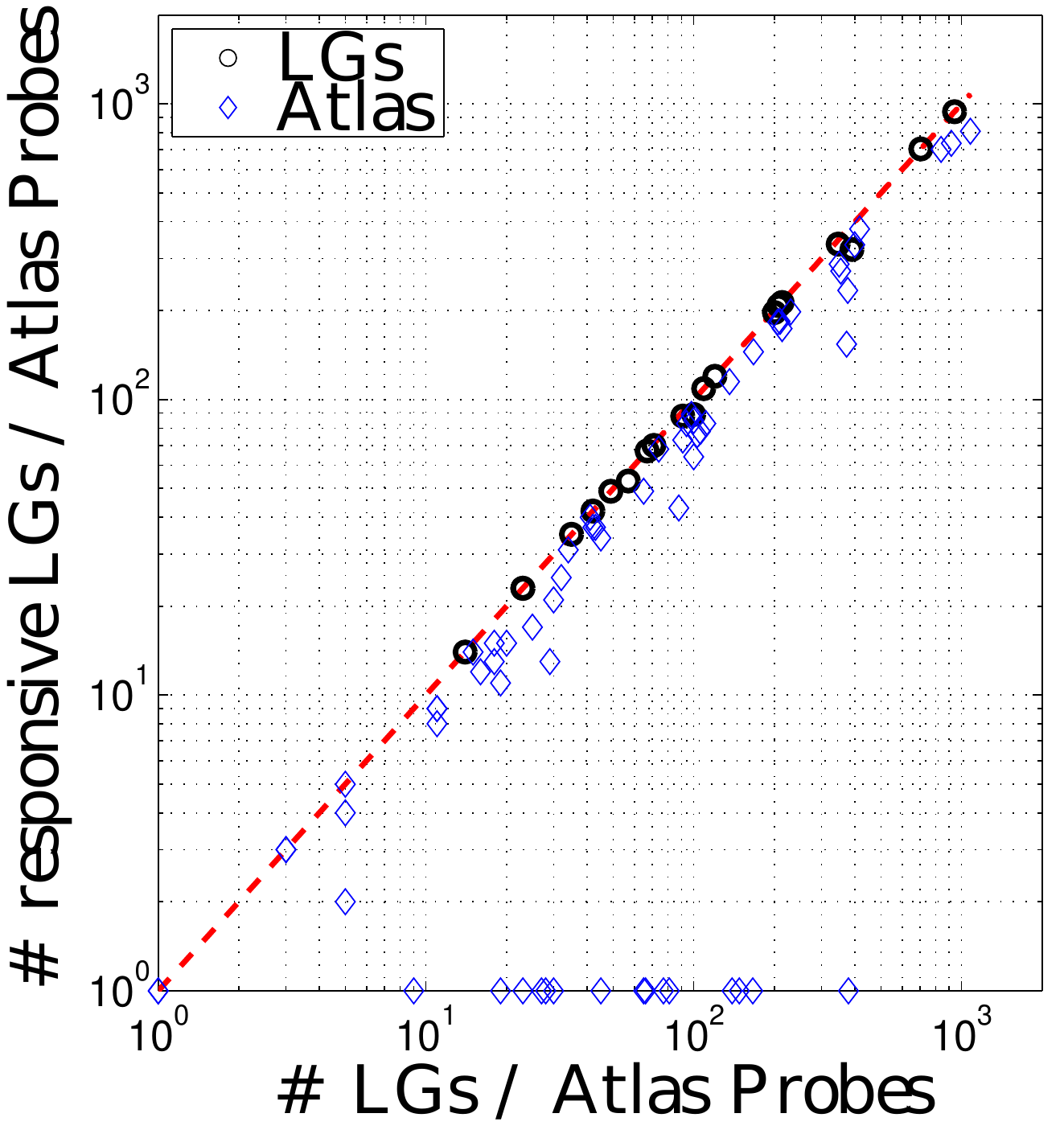}
		\label{fig:vp-responses}
	}
    \hspace{0.1em}
	\subfloat[ECDF of the the minimum RTT for each responsive IXP peering interface.]{
		\includegraphics[clip,width=4cm,valign=t]{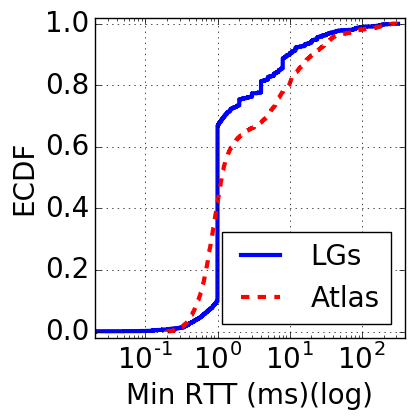}
		\label{fig:ping-rtts}
	}
    \hspace{0.1em}
 	\subfloat[Inference result compared to number of feasible facilities and $RTT_{min}$ for each interface.]{
 		\includegraphics[clip,width=4cm,valign=t]{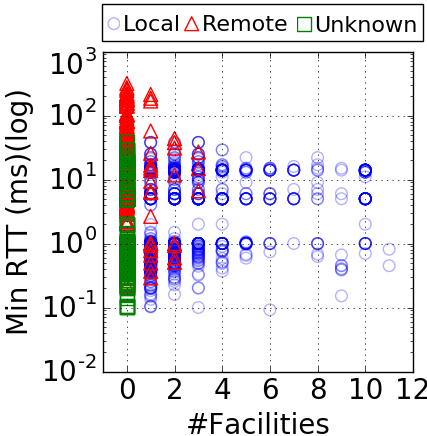}
 		\label{fig:feasible-vs-rtt}
 	}
    \hspace{0.1em}
 	\subfloat[Multi-IXP router types compared to number of next-hop IXPs.]{
 		\includegraphics[clip,width=4cm,valign=t]{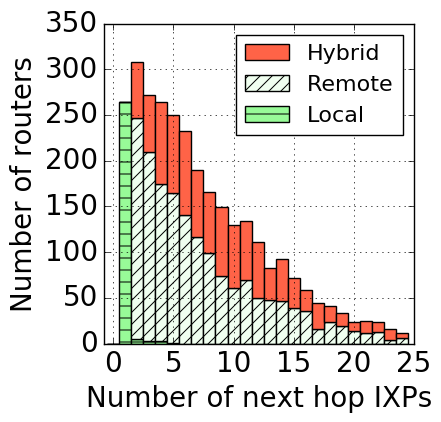}
 		\label{fig:multirouter-type}
 	}
	\vspace{-3mm}
	\caption{Measurement results for RTTs, feasible facilities and multi-IXP routers. }
	\label{fig:lg-ping-rtts}
    \vspace{-1.2em}
\end{figure*}

Here, we apply our inference methodology on the 30 largest IXPs in our dataset, step by step (Section~\ref{subsec:step-inf-wild}). Having inferred RP at IXPs, we investigate some relevant use cases. Indicatively, we focus on RP features in Section~\ref{subsec:analysis-rp-features}. We further study aspects of the evolution of the RP ecosystem over time (Section~\ref{subsec:analysis-trends}), as well as routing implications involving a large IXP (Section~\ref{subsec:analysis-rp-impl}).

\subsection{Application of Step-wise Inference}
\label{subsec:step-inf-wild}

\begin{figure*}[!ht]
\centering
\subfloat[Contribution of each inference step per IXP.]{
		\includegraphics[clip,valign=t,width=7.5cm]{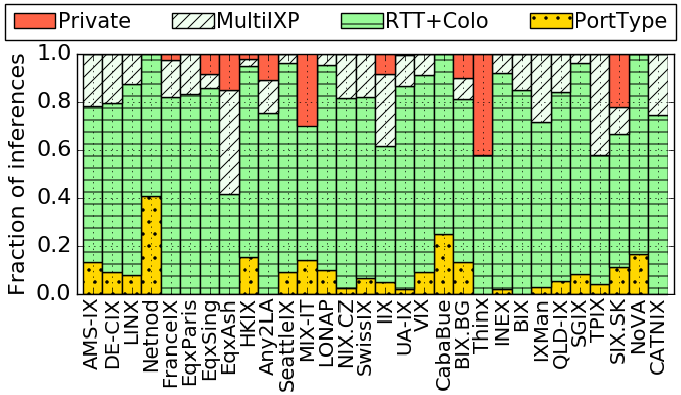}
		\label{fig:inference-heuristic-contribution}
	}
    \hspace{0.1em}
	\subfloat[Inferences per IXP.]{
		\includegraphics[clip,width=9.5cm,valign=t]{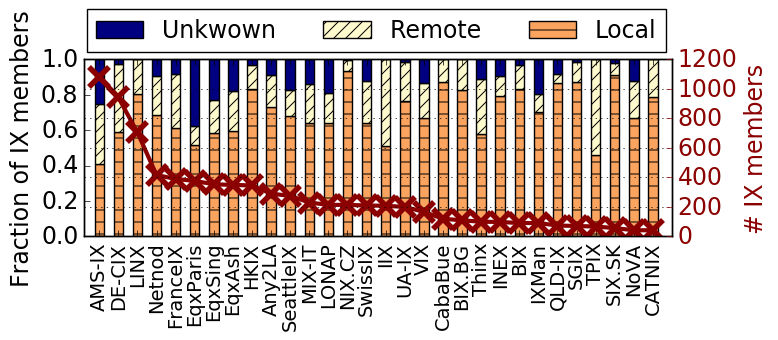}
		\label{fig:inferences_per_ixp}
	}
    \hspace{0.1em}

	\vspace{-3mm}
	\caption{Inference results for the 30 largest IXPs with LG/Atlas VPs.}
	\label{fig:inferences}
    \vspace{-1em}
\end{figure*}

\myitem{Step 1.} We first infer the IXP members that reach the IXP through resellers, by comparing the port capacities of each member against the minimum physical port capacity offered by that IXP. For some IXPs (see Fig.~\ref{fig:inference-heuristic-contribution}), such as France-IX, which cooperates with more than $20$ resellers~\cite{franceixresellers}, 40\% of the inferences can be made by using only port capacity information. However, for other IXPs that do not allow port reselling (\eg HKIX), this step fails to make any inference. On average, this step contributes for approx. 10\% of the total inferences (see column COV in Table~\ref{tab:validation}).

\myitem{Step 2.} We then execute a ping measurement campaign between 7-9~Apr.~2018 from each LG and Atlas VP, to the peering interfaces of the IXP that hosts the VP. LGs achieve high response rates (Fig.~\ref{fig:vp-responses}) due to being directly attached to the IXP peering LAN. In contrast, 50 of the 66 Atlas probes are colocated within an IXP facility, but are not inside the IXP's LAN. Therefore, pings from them to IXP LAN IP addresses are more likely to fail for various reasons~\cite{mason2001cisco}.
14 of the Atlas probes do not receive any ping response.

Figure~\ref{fig:ping-rtts} shows the $RTT_{min}$  distributions between VPs (LGs and Atlas probes) and IXP interfaces. 75\% of the IXP interfaces are within $2ms$ from the respective VP. \textbf{More than 20\% of the interfaces have $\mathbf{RTT_{min} > 10ms}$, a 2-fold increase since 2014}~\cite{castro_dataset, castro2014remote}.

However, we found Atlas probes with consistently inflated RTT values\footnote{Atlas probes can yield measurement errors~\cite{holterbach2015quantifying}; in our campaign, we account for non-persistent inflation by considering minimum RTTs over time.}.
Such probes may be deployed in the IXP's management LAN which may not be in the IXP's facilities, but still abide to the \textit{TTL match} filter (see Section~\ref{subsec:rtt-limitations}) which is set to $TTL_{max} - 1$ for Atlas probes. Thus, we discard probes that have $RTT_{min} \geq 1ms$ between the probe and the IXP's route server. This filter removes another $21$ Atlas probes from the set of usable VPs. Also, note that a large number of minimum RTTs obtained from LGs are exactly $1ms$, which happens because many LGs round up the RTT value to the nearest integer. For such LGs we calculate the $d_{min}$ distance between the IXP interface and the VP assuming $RTT'_{min} = RTT_{min} - 1$, and we use the rounded-up $RTT_{min}$ to calculate the corresponding $d_{max}$ distance. 

Table~\ref{tab:ping-iface-stats} provides the statistics of the queried interfaces that were used for our inferences after filtering out the unusable VPs. 

\myitem{Step 3.} We calculate the feasible IXP and AS facilities for each peering interface, based on the measured $RTT_{min}$, and infer the interfaces as local, remote or unknown, based on the combined latency and colocation information. Figure~\ref{fig:feasible-vs-rtt} shows the $RTT_{min}$ for each IXP interface versus the number of feasible facilities. Each $(RTT_{min}, \#facilities)$ data point is tagged with its inferred peering type. \textbf{94\% of the remote interfaces have no feasible common facility with the IXP} (which further validates the colocation ``principle''), while for 6\% we have at least one feasible facility. Drilling down on this 6\%, 40\% of the involved interfaces exhibit $RTT_{min} > 2ms$, indicating spurious colocation information. Moreover, 5\% of them are in a facility within the same metro area as the IXP VP but not affiliated with the IXP, while the rest are cases of IXP members colocated with the IXP but connecting through a reseller via a low-capacity virtual port (inferred at Step 1).

\myitem{Step 4.} For the \textit{unknown} interfaces of Step 3, we investigate if they are part of multi-IXP routers. Figure~\ref{fig:multirouter-type} shows the number (per inferred type) of IXP routers compared to the number of IXPs with which they are connected (next-hop IXPs). Surprisingly, we find that 20\% of the \textit{unknown} interfaces and $\sim$\textbf{80\% of the corresponding routers have multiple IXP connections, with 25\% of them connecting to more than 10 IXPs}. 
This result highlights that the AS-level and IXP-level peering diversity of such IXP peers are misleading indicators of their resilience, since \textbf{all of their interconnections depend on the same physical equipment} (\ie{the multi-IXP router}). We further observe that cases of remote multi-IXP routers are more prevalent than hybrid ones.

\begin{table}[!b]
\centering
\small
\caption{Statistics of interfaces involved in the ping campaign. For our measurements we used the 30 largest IXPs with usable VPs.}
\vspace{-1em}
\label{tab:ping-iface-stats}
\begin{tabular}{c|c|c|c|c|c}
\hline
\multirow{2}{*}{\textbf{VP Type}}& 
\multirow{2}{*}{\textbf{\# VPs}} & 
\multicolumn{2}{|c|}{\textbf{\# Interfaces}}& \multirow{2}{*}{\textbf{\# Members}}& \multirow{2}{*}{\textbf{\# IXPs}}\\
&&Queried& Resp. (Fig.~\ref{fig:vp-responses})&&\\
\hline
\hline
LG&23&3,806&3,617 (95\%)&2,347&18\\
\hline
Atlas&22&6,457&4,861 (75\%)&4,097&22\\
\hline
\hline
Total&45&10,578&7,738 (73\%)&6,444&30\\
\hline
\end{tabular}
\end{table}
 
\myitem{Step 5.} Finally, for the remaining unknown interfaces we infer locality or remoteness based on their private connectivity. As shown in Fig.~\ref{fig:inference-heuristic-contribution}, we had to apply this heuristic only for 11 of the top 30 IXPs, because previous steps did not manage to successfully infer some of the IXP interfaces of these IXPs as remote or local.

\myitem{Overall.} In total, the contribution (in terms of fraction of inferences) of each step of the methodology is shown in Fig.~\ref{fig:inference-heuristic-contribution}. Steps 2 ($RTT+colo$) and 3 (multi-IXP routers) account for the majority of the inferences. Moreover, Fig.~\ref{fig:inferences_per_ixp} shows the final inference results for the top 30 IXPs. Overall,\textbf{ we find 28\% of all the IXP interfaces for which we made an inference to be remote. Also, for 90\% of the IXPs, it holds that more than 10\% of their members are remote peers.} Finally, we find that for the two largest IXPs (DE-CIX and AMS-IX) almost 40\% of their members are remote.

\subsection{Features of Remote Peers}
\label{subsec:analysis-rp-features}

Having inferred remote and local peers per IXP, we proceed to investigate what are the features of remote peers and if/how they differentiate from local peers. We examine 3 features for each IXP member: (i) the size of its customer cone, as reported by CAIDA~\cite{customer_cones}, (ii) its traffic levels and countries of presence as reported by PDB~\cite{peeringdb}, and (iii) the user population it serves, as reported by APNIC~\cite{apnic-ipv6}. We classify an IXP member network as follows: ``remote'' if it has only remote connections; ``local'' if it has only local connections; ``hybrid'' if it has both types (in the same or multiple IXPs).
Out of 2959 total inferred AS-peers in 30 IXPs, we find that 63.7\% are local, 23.4\% are remote and 12.9\% are hybrid.

\begin{figure}[t]
\subfloat[Customer cones of inferred local (blue), remote (red), hybrid (green) IXP members.]
    {
		\includegraphics[width=0.5\columnwidth,valign=t]{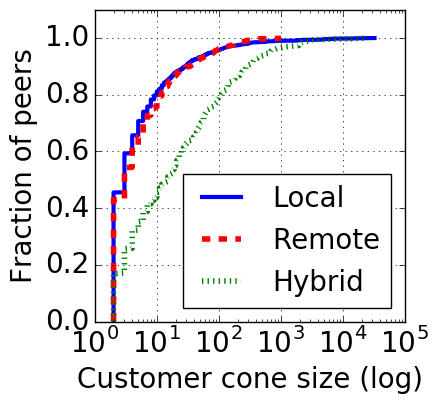}
	    \label{fig:customer-cones}
    }
    \hspace{0.02cm}
    	\subfloat[Traffic levels of inferred local, remote and hybrid IXP members.]{
		\includegraphics[clip,height=3.9cm,valign=t]{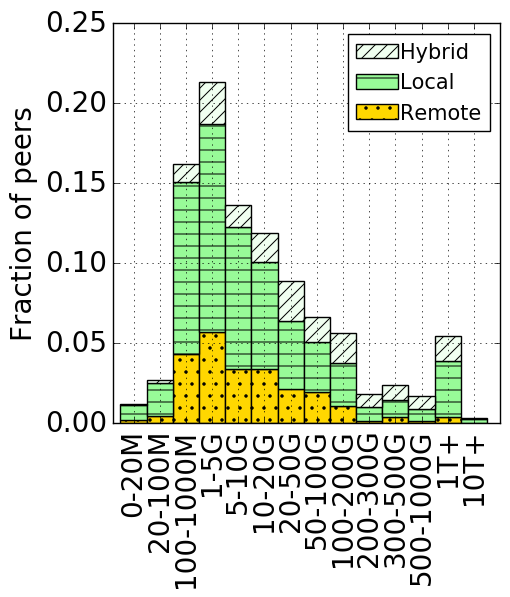}
		\label{fig:traffic-levels}
	}
    \vspace{-3mm}
    \caption{Features of all inferred IXP members.}
    \label{fig:feature-peers}
	\vspace{-1.2em}
\end{figure}

In Fig.~\ref{fig:customer-cones} we show the fractions of remote, local and hybrid IXP members with respect to the size of their customer cone. We observe that remote peers (red line) have quite similar patterns with the local ones (blue line). In fact, whether a network chooses to engage in local or remote peering (which is a matter of network design) at an IXP is not reflected on the size of its customer cone. This is probably due to the fact that both practices achieve similar Internet reachability to/from the local/remote peer's customers. Interestingly enough, member ASes that are local peers in some IXPs and remote in others tend to have one order of magnitude larger customer cones than the other cases. This is because hybrid IXP members are usually large ISPs that have diverse peering policies over large geographical areas, engaging both in local and remote peering depending on their business needs per market segment. Note that the insights pertaining to the customer cones of local, remote and hybrid peers are also reflected in the estimated user populations by APNIC, as expected (results omitted for brevity). 

Regarding the country distribution of the IXP members, we found that most local (13.86\%) and hybrid (11.04\%) peers are headquartered in GB, while PL seems to host the most remote peers (12.88\%). 

With respect to the traffic levels associated with each network\footnote{In Figure~\ref{fig:traffic-levels}, we refer to the aggregate --self-reported via PeeringDB-- traffic levels exchanged by the network themselves and not their peering connections.}, as shown in Fig.~\ref{fig:traffic-levels}, the observed pattern seems to comply with the insights related to the cones and user populations of the IXP members. The distributions of the traffic levels for remote and local peers are similar (albeit with the fraction of local peers per traffic level being larger as expected), while hybrid peers seem to be present also at very high traffic levels, together with locals. It is also interesting that networks with vastly different traffic levels (ranging from 100s of Mbits to 100s of Gbits) engage in RP practices.

\subsection{RP Evolution}
\label{subsec:analysis-trends}

\begin{figure}[t]
\centering
\subfloat[The increase of remote peers is 2x faster compared to the increase of local peers.]{
\includegraphics[clip,valign=t,width=4cm]{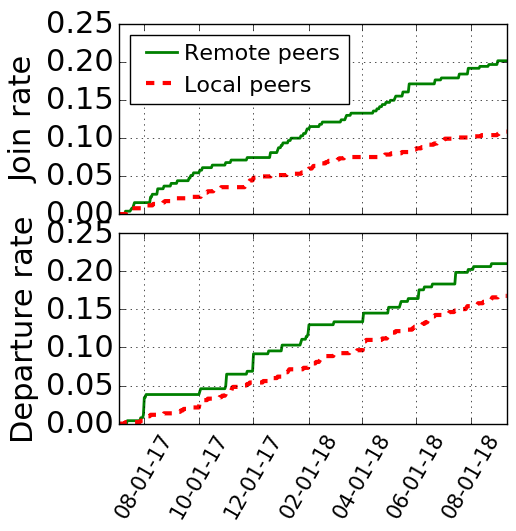}
		\label{fig:historical-trends}
	}
    \hspace{0.1em}
	\subfloat[Comparison of ping and traceroute RTTs for LINX LON peering interfaces.]{
\includegraphics[clip,width=4cm,valign=t]{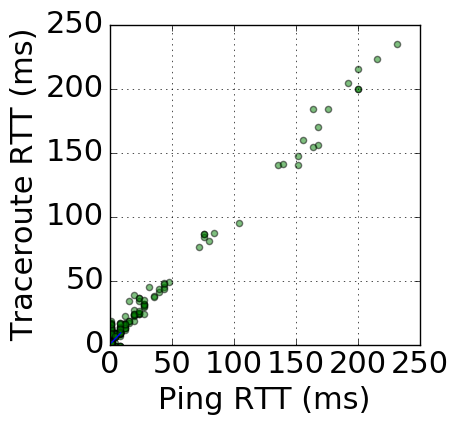}
		\label{fig:remote-local}
	}
    \hspace{0.1em}

	\vspace{-3mm}
	\caption{Analysis of archived RTT data.}
	\label{fig:inferences}
    \vspace{-1.2em}
\end{figure}

To understand aspects of the evolution of RP over time, we collect (i) daily RTT measurements (pings) from available LG VPs in 5 IXPs (LINX, HKIX, LONAP, THINX and UAIX), (ii) PDB dumps, and (iii) Atlas traceroutes between
2017/07/04 - 2018/09/10, and we use them to infer remote and local peers across time. Based on this information, we can calculate aggregate growth (\ie{a new member joins an IXP}) and departure (\ie{an old member leaves an IXP}) rates \emph{per peering type}.
We observe that \textbf{the number of remote peers grows twice as fast as the number of local peers}, indicating that today, remote peers are the primary drivers of IXP growth (Fig.~\ref{fig:historical-trends}). These results are
confirmed by IXP annual reports from some of the largest IXPs (AMS-IX, DE-CIX, France-IX)~\cite{amsix-annual-report,decix-annual-report,franceix-annual-report},
indicating that IXPs that already service the majority of local networks in their respective country-level peering ecosystems,
seek to expand their market pool by attracting remote peers. However, remote peers also exhibit higher ($+25\%$) departure rates than local ones; reseller customers do not commit substantial resources to establish their IXP connectivity (\eg{routing equipment at the IXP}), therefore it might be easier for them to terminate it. For the same time period we  also found 18 cases of peers that switched from remote to local interconnections.

\subsection{RP Routing Implications}
\label{subsec:analysis-rp-impl}

Here, as another use case demonstrating the applicability of our inference methodology, we investigate the interplay between remote peering and Internet routing. Specifically, we consider the DE-CIX Frankfurt (FRA) IXP, and examine the routing behavior between its 314 remote members (as inferred by our methodology) and any other of its 781 (local or remote) members (available at the time of measurement). Let $AS_{R}$ be a remote member of DE-CIX FRA, and $AS_{x}$ another DE-CIX FRA member (remote or local; $AS_{R}\neq{AS_{x}}$), which peers in at least one more common IXP with $AS_{R}$. We are interested in \emph{circuitous paths} that start at $AS_{R}$ and end at $AS_{x}$, which we find with the following process. (i) We randomly choose maximum 5 available (up and running) RIPE Atlas probes within $AS_{R}$. (ii) We extract the routed prefixes that $AS_{x}$ advertises via BGP, using the RIPEstat service~\cite{ripestatservice}. (iii) We select the first IP address (.1) of a randomly chosen prefix among these prefixes. (iv) We run traceroutes from the chosen probes in $AS_{R}$ towards the selected IP address of $AS_{x}$. (v) We extract all traceroute paths involving an IXP crossing (see Section~\ref{subsec:traixroute}), either over DE-CIX FRA or another common IXP.

We analyze the results for all possible \{$AS_{R}$, $AS_{x}$\} pairs ($\sim245k$~in total). We identify 5941 IXP crossings involving $AS_{R}$ and $AS_{x}$ as the two peering IXP members. As described above, these crossings involve either DE-CIX FRA or another IXP where both ASes peer. $AS_{R}$ and $AS_{x}$ are also the source and destination of the traceroute(s), respectively. In the majority of the cases (66\%), we observe that the routing decision of $AS_{R}$ seems to comply with an expected hot-potato exit strategy~\cite{caesar2005bgp,teixeira2004dynamics}, \ie{the IXP involved in the crossing is the closest one to $AS_{R}$ among the IXPs where both $AS_{R}$ and $AS_{x}$ are present.} Interestingly enough, on the one hand, we identify cases (18\%) where traffic is exchanged via the RP interconnection of $AS_{R}$ at DE-CIX FRA, while there exists another common IXP that is closer to $AS_{R}$. By using this closer IXP, instead of the RP in DE-CIX FRA, $AS_{R}$ could offload traffic 100s of km closer to its network. On the other hand, there are cases (16\%) where the two peers use another (local or remote) peering link (\ie{not over DE-CIX FRA}) to exchange traffic, while the facilities of DE-CIX FRA are closer to the $AS_{R}$. In the latter cases, $AS_{R}$ could use the RP over DE-CIX FRA to offload traffic hundreds of km closer to its network.
 
The reason why in some cases these networks do not make a ``seemingly better'' (latency-wise) routing decision has to do with their own policies, which are not typically known. Note also that both routing options (over remote peering in DE-CIX or local/remote peering in another IXP), are over peer-to-peer links via IXPs. Therefore, we cannot distinguish routing preferences based on coarse AS-relationships~\cite{luckie2013relationships} (customer \textit{vs.} peer \textit{vs.} provider); this would require taking into account also additional features, such as BGP communities~\cite{giotsas2014inferring}, which is the subject of future work.

\section{Discussion \& Insights}
\label{sec:discussion}

\myitem{Ubiquity and growth.} We found that RP becomes an increasingly popular practice and is almost ubiquitous in the global IXP ecosystem {(Fig.~\ref{fig:inferences_per_ixp})}. For instance, in AMS-IX, $90\%$ of new customers come through reseller programs~\cite{euroix2016}. Exceptions are IXPs that do not support port reselling, but even they facilitate RP for physically distant members, e.g., over L2 carriers. It is worth mentioning that in the past, one of the reasons why local IXP traffic remained local, was because distance meant cost. In light of significantly reduced transport costs, that is no longer the case. Indeed, the largest IXPs have more distant members compared to the average. This is an example of a network effect; the more members an IXP has, the more valuable that IXP is to networks~\cite{Norton-remote-peering}. We observe that most new members at the largest IXPs are remote {(Fig.~\ref{fig:historical-trends})}. Interestingly, even smaller IXPs exhibit growing tendencies in terms of RP.

\myitem{Implications.}
There are RP cases that have a clear impact on routing paths, and thus, on performance and resiliency.
We find evidence that RPs support suboptimal routing choices and introduce latency penalties (Section~\ref{subsec:analysis-rp-impl}). In fact, in many cases, exchanging traffic at an IXP where both traffic source and destination are colocated as members, would be more beneficial for performance {(\eg{lower latency; Fig.~\ref{fig:ground-truth-rtt}})}. For ASes with a global footprint the lack of visibility in whether a peer is local or remote makes their traffic engineering considerably harder. In particular, anycast routing employed by CDNs is affected by RP practices that drive traffic away from the intended load-balancing center, i.e., the IXP itself. In contrast, we also find cases where RP can improve performance by offering better routing choices to a broader set of networks (Section~\ref{subsec:analysis-rp-impl}).

In terms of resilience, there are potential issues with RP setups.
While an extensive investigation of these issues is the subject of future work, here we reason about some obvious resilience implications. Multiple peers connect via the same reseller's physical port to one IXP {(see Fig.~\ref{fig:portspeeds-gt}}); one remote peer may connect (through different resellers) to multiple IXPs  with just one router {(see Fig.~\ref{fig:multirouter-type})}. In these cases, an outage in a single IXP or IXP switch port: (i) propagates much further than the metropolitan area of the IXP over the RP link, and (ii) may affect several IXP members at once. Even if these members have backup paths/links that are activated upon such a cascading outage (\eg over their transit providers), there is some unavoidable delay and packet loss until BGP routing converges on the backup paths~\cite{giotsas2017detecting}; even after this happens, routing may be sub-optimal in terms of packet-level end-to-end delay.

\myitem{The IXP's point of view.} IXPs themselves do not discriminate members as local or remote; they are simply interested in (i) attracting as many ASes as possible (potentially expanding in multiple geographical regions and colocation facilities or via RP), and offering (ii) short paths and low latencies, (iii) high throughput, and (iv) additional services. However, they are aware of which customers are ``virtual'' {(Section~\ref{subsec:groundtruth_remote_members})}, \ie{connected on a virtual port offered by a port reseller}, since they need to terminate the inner VLAN and configure the IXP-end of the virtual port. For IXPs, RP inference (uncovering also distant, non-virtual IXP peers{; see Steps 2-5 in Section~\ref{subsec:inference_algorithm}}) is interesting for two main reasons: (a) to overcome the local saturation and find new attractive markets/locations for expansion, and (b) to offer to their customers transparency as to who is local and who is not. Finally, we note emerging RP flavors. For instance, NL-IX is a reseller for AMS-IX; DE-CIX offers the \textit{GlobePeer}~{\cite{decix-globepeer}} product that allows an IXP member to acquire access to all DE-CIX peers irrespective of their location.

\section{Future Research Directions}
\label{sec:future-work}

Being equipped with a methodology that reliably infers RP, we identify the following directions for future work in this field.

\myitem{Traffic Analysis.}
Knowing which IXP peering interconnections are remote/local, a natural follow-up step is to investigate the importance of RP in terms of the actual traffic flowing over the switching infrastructure of an IXP. To achieve this, we need datasets containing the traffic levels of remote and local IXP peering connections.

\myitem{Beyond Pings.} 
Measuring RTTs using pings from VPs within the IXP suffers from limited and unstable VPs. However, traceroutes from VPs anywhere in the world, can provide an additional source of useful RTT measurements that cover much more in space and time than pings. In fact, the difference of the RTTs between the VP and the consecutive IP interfaces involved in a --potential-- IXP crossing, as observed in a traceroute path, can provide an indication of the delay between the associated IXP peers. Fig.~\ref{fig:remote-local} shows the RTTs between the LINX LON LG and the interfaces of the members of LINX, using traceroutes (see Section~\ref{subsec:msm_sources}) and pings; we observe that the RTT patterns are close, supporting such an approach. However, traceroute-based approaches come with their own set of challenges such as asymmetric paths, load-balancing artifacts, ICMP rate-limits and heterogeneous opaque layer-2 connectivity mechanisms~\cite{luckie2014challenges, steenbergen2009practical}. We plan to investigate inference approaches that are robust against such artifacts and enable us to scale up our methodology, decoupling it from ping-based measurements.

\myitem{Longitudinal Study.}
In Section~\ref{subsec:analysis-trends} we analyzed the RP growth of 5 IXPs over a period spanning more than a year. Understanding whether our observations represent an actual trend, and not just recent developments in the IXP ecosystem, requires digging deeper into history. Since daily RTT measurements (pings) from LGs in IXPs are not available for all IXPs (e.g., during the time-frame of Section~\ref{subsec:analysis-trends}), we aim to apply a traceroute-based methodology to perform an extensive analysis in space (more IXPs) and time (years).

\myitem{Other Implications/Trade-offs.}
Evaluating the impact of RP and routing policies (Section~\ref{subsec:analysis-rp-impl}) on the performance of CDNs and anycast services might be of interest to the community. RP is associated to implications for performance, resilience/reliability, and security, and it comes with certain trade-offs (e.g., debugging is more complex when third-party layer-2 infrastructures are involved). Follow-up work could focus on assessing such trade-offs and comparing RP to more traditional connectivity practices, such as classic transit.

\vspace{-0.3em}
\section{Conclusion}
\label{sec:conclusions}

In this work, we introduce, validate, and apply a methodology that can infer remote and local peers at IXPs with high accuracy and coverage. Our methodology is built upon the observation that RP is not driven only by technical factors, but is actually a business decision guided by economic considerations. In particular, taking into account port capacities, colocation strategies, multi-IXP peerings as well as latencies and private connectivity practices, we achieve very high accuracy (95\%) and coverage (93\%) levels, outperforming the state-of-the-art by $+18\%$ and $+9\%$, respectively. At the same time we reduce almost 4 times the false positive and negative rates. The primary objective of this approach is to enable IXPs and existing or new potential IXP members to understand which peers of an IXP are \emph{physically} local, allowing for better-informed peering and routing decisions. Moreover, by equipping researchers with a reliable inference methodology, we enable the in-depth investigation of multiple facets of the  peering ecosystem, such as the detection of routing inefficiencies that undermine the resilience and performance of traffic exchange. In our measurement-based study of 30 of the largest IXPs worldwide, we found that more than 90\% of them have more than 10\% of their members as remote peers. Strikingly, for large IXPs, this share may exceed 40\%. The number of remote peers grows twice as fast compared to local peers, driving the IXP growth. The remote peers show similar patterns with local peers in terms of customer cones, user populations and aggregated traffic levels, indicating that remote peering is widely adopted practice across networks. Moreover, we observe that several remote peer routers are connected to more than 10 IXPs, while we also find evidence of hybrid (remote \& local) IXP peering interconnections on the same router, with profound implications for routing resilience.

\myitem{Prototype and Portal.}
To automate our remote peering inference methodology and make our results publicly accessible to the community, we have implemented a web portal at~\cite{remote-peering-portal}, through which we publish monthly snapshots of our inferences, and visualize the geographical footprint of IXPs and their connected members.

\myitem{Acknowledgements.} We thank the anonymous reviewers and Walter Willinger for their constructive feedback, and Eleni Fragkiadaki for her technical insights. This work has been funded by the European Research Council grant agreement no. 338402.

\newpage
\bibliographystyle{acm}
\bibliography{references}

\begin{thebibliography}{10}

\bibitem{amsixam_members}
{AMS-IX Amsterdam: connected networks}.
\newblock \url{https://ams-ix.net/connected_parties}.
\newblock Accessed: 13.05.2018.

\bibitem{amsixam_pricing}
{AMS-IX Amsterdam: services and pricing}.
\newblock \url{https://ams-ix.net/services-pricing/pricing}.
\newblock Accessed: 13.05.2018.

\bibitem{amsixam_stats}
{AMS-IX Amsterdam: traffic statistics}.
\newblock \url{https://ams-ix.net/technical/statistics}.
\newblock Accessed: 13.05.2018.

\bibitem{amsix-easyaccess}
{AMS-IX EasyAccess Service}.
\newblock \url{https://ams-ix.net/services-pricing/easyaccess}.
\newblock Accessed: 13.05.2018.

\bibitem{customer_cones}
{CAIDA AS Relationships}.
\newblock \url{http://www.caida.org/data/as-relationships}.

\bibitem{coresite_carrier_list}
{Coresite Carrier List}.
\newblock \url{
  https://www.coresite.com/resources/resource-library/additional/carrier-list}.
\newblock Accessed: 24.09.2018.

\bibitem{decixfr_members}
{DE-CIX Frankfurt connected networks}.
\newblock
  \url{https://www.de-cix.net/en/locations/germany/frankfurt/connected-networks}.
\newblock Accessed: 13.05.2018.

\bibitem{decixfr_stats}
{DE-CIX Frankfurt traffic statistics}.
\newblock
  \url{https://www.de-cix.net/en/locations/germany/frankfurt/statistics}.
\newblock Accessed: 13.05.2018.

\bibitem{decix-globepeer}
{DE-CIX GlobePEER Remote Service}.
\newblock
  \url{https://www.de-cix.net/en/de-cix-service-world/globepeer-remote}.
\newblock Accessed: 13.05.2018.

\bibitem{franceixresellers}
{France-IX Resellers List}.
\newblock \url{https://www.franceix.net/en/members-resellers/resellers}.
\newblock Accessed: 13.05.2018.

\bibitem{he_dataset}
{Hurricane Electric, Internet Exchange Report.}
\newblock \url{https://bgp.he.net/report/exchanges}.
\newblock Accessed: 27.03.2018.

\bibitem{inflect}
{Inflect: Find the right data center}.
\newblock \url{https://inflect.com}.
\newblock Accessed: 13.05.2018.

\bibitem{ixreach_rp_service}
{IX Reach Remote Peering Service}.
\newblock \url{http://ixreach.com/services/remote-peering}.
\newblock Accessed: 13.05.2018.

\bibitem{nlix}
{NL-IX: The Interconnect Exchange}.
\newblock \url{https://www.nl-ix.net}.
\newblock Accessed: 13.05.2018.

\bibitem{pch_dataset}
{Packet Clearing House, Internet Exchange Directory}.
\newblock \url{https://prefix.pch.net/applications/ixpdir/menu_download.php}.
\newblock Accessed: 30.04.2018.

\bibitem{peeringdb}
{PeeringDB}.
\newblock \url{https://www.peeringdb.com}.
\newblock Accessed: 30.04.2018.

\bibitem{remote-peering-portal}
{Remote IXP Peering Portal}.
\newblock \url{http://remote-ixp-peering.net}.

\bibitem{retn_rp_service}
{RETN Remote Peering Service}.
\newblock \url{http://retn.net/services/remote-ix}.
\newblock Accessed: 13.05.2018.

\bibitem{RipeAtlas}
{RIPE Atlas - Measurements}.
\newblock \url{https://atlas.ripe.net/about/measurements}.
\newblock Accessed: 13.05.2018.

\bibitem{ripestatservice}
{RIPEStat Service}.
\newblock \url{https://stat.ripe.net}.
\newblock Accessed: 13.05.2018.

\bibitem{traixroutegit}
{traIXroute - Source Code}.
\newblock \url{https://github.com/gnomikos/traIXroute}.

\bibitem{itu1731}
{ITU-T Y.1731 Performance Monitoring In a Service Provider Network}.
\newblock
  \url{https://www.cisco.com/c/en/us/td/docs/ios/cether/configuration/guide/ce_y1731-perfmon.html},
  Mar 2011.
\newblock Accessed: 13.05.2018.

\bibitem{euroix2016}
{Remote Peering Panel Discussion, \nth{29} Euro-IX Forum, Krakow, Poland}.
\newblock \url{https://www.euro-ix.net/en/events/fora/29th-euro-ix-forum}, Nov
  2016.

\bibitem{NetIXMap}
{NET-IX Network Map}.
\newblock \url{https://www.netix.net/network_map}, Apr 2018.

\bibitem{ager2012anatomy}
{\sc Ager, B., Chatzis, N., Feldmann, A., Sarrar, N., Uhlig, S., and Willinger,
  W.}
\newblock {Anatomy of a large European IXP}.
\newblock {\em ACM SIGCOMM CCR 42}, 4 (2012), 163--174.

\bibitem{ahmed2017peering}
{\sc Ahmed, A., Shafiq, Z., Bedi, H., and Khakpour, A.}
\newblock Peering vs. transit: Performance comparison of peering and transit
  interconnections.
\newblock In {\em Proc. of IEEE ICNP\/} (2017).

\bibitem{amsix-annual-report}
{\sc {AMS-IX}}.
\newblock {AMS-IX 2016 Annual report}.
\newblock \url{https://ams-ix.net/annual_report/2016}.
\newblock Accessed: 05.09.2018.

\bibitem{apnic-ipv6}
{\sc {APNIC}}.
\newblock {IPv6 Measurement Campaign}.
\newblock \url{https://stats.labs.apnic.net/v6pop}.
\newblock Measurement Methodology: \url{https://labs.apnic.net/measureipv6},
  Dataset collected on: 22.05.2017.

\bibitem{augustin2009ixps}
{\sc Augustin, B., Krishnamurthy, B., and Willinger, W.}
\newblock {IXPs: mapped?}
\newblock In {\em Proc. of ACM IMC\/} (2009).

\bibitem{bottger2017hypergiant}
{\sc B{\"o}ttger, T., Cuadrado, F., Tyson, G., Castro, I., and Uhlig, S.}
\newblock {A Hypergiant's View of the Internet}.
\newblock {\em ACM SIGCOMM CCR 47}, 1 (2017).

\bibitem{caesar2005bgp}
{\sc Caesar, M., and Rexford, J.}
\newblock {BGP routing policies in ISP networks}.
\newblock {\em IEEE network 19}, 6 (2005), 5--11.

\bibitem{iffinder}
{\sc CAIDA}.
\newblock {iffinder}.
\newblock \url{http://www.caida.org/tools/measurement/iffinder}.

\bibitem{kapar}
{\sc CAIDA}.
\newblock {kapar}.
\newblock \url{http://www.caida.org/tools/measurement/kapar}.

\bibitem{caidapfxtoas}
{\sc CAIDA}.
\newblock {Routeviews prefix2as Dataset}.
\newblock \url{http://data.caida.org/datasets/routing/routeviews-prefix2as}.

\bibitem{castro_dataset}
{\sc Castro, I., Cardona, J.~C., Gorinsky, S., and Francois, P.}
\newblock {Remote Peering Data}.
\newblock \url{https://svnext.networks.imdea.org/repos/RemotePeering}.
\newblock Accessed: 13.05.2018.

\bibitem{castro2014remote}
{\sc Castro, I., Cardona, J.~C., Gorinsky, S., and Francois, P.}
\newblock {Remote peering: More peering without internet flattening}.
\newblock In {\em Proc. of ACM CoNEXT\/} (2014).

\bibitem{chatzis2013benefits}
{\sc Chatzis, N., Smaragdakis, G., B{\"o}ttger, J., Krenc, T., and Feldmann,
  A.}
\newblock {On the benefits of using a large IXP as an Internet vantage point}.
\newblock In {\em Proc. of ACM IMC\/} (2013).

\bibitem{chatzisimportance}
{\sc Chatzis, N., Smaragdakis, G., Feldmann, A., and Willinger, W.}
\newblock {On the Importance of Internet eXchange Points for Today's Internet
  Ecosystem}.
\newblock {\em ACM SIGCOMM CCR\/} (2013).

\bibitem{chatzis2013there}
{\sc Chatzis, N., Smaragdakis, G., Feldmann, A., and Willinger, W.}
\newblock {There is more to IXPs than meets the eye}.
\newblock {\em ACM SIGCOMM CCR 43}, 5 (2013), 19--28.

\bibitem{chatzis2015quo}
{\sc Chatzis, N., Smaragdakis, G., Feldmann, A., and Willinger, W.}
\newblock {Quo vadis Open-IX?}
\newblock {\em ACM SIGCOMM CCR 45}, 1 (2015), 12--18.

\bibitem{chiu2015we}
{\sc Chiu, Y.-C., Schlinker, B., Radhakrishnan, A.~B., Katz-Bassett, E., and
  Govindan, R.}
\newblock {Are we one hop away from a better internet?}
\newblock In {\em Proc. of ACM IMC\/} (2015).

\bibitem{decix-annual-report}
{\sc {DEC-IX}}.
\newblock {DE-CIX Annual Report 2016}.
\newblock \url{https://goo.gl/qwCM23}.
\newblock Accessed: 05.09.2018.

\bibitem{fanou2017investigating}
{\sc Fanou, R., Valera, F., and Dhamdhere, A.}
\newblock {Investigating the Causes of Congestion on the African IXP
  substrate}.
\newblock In {\em Proc. of ACM IMC\/} (2017).

\bibitem{franceix-annual-report}
{\sc {FRANCE-IX}}.
\newblock {France-IX Annual Report 2017}.
\newblock \url{https://www.franceix.net/annual-report-2017}.
\newblock Accessed: 05.09.2018.

\bibitem{giotsas2016periscope}
{\sc Giotsas, V., Dhamdhere, A., and Claffy, K.~C.}
\newblock {Periscope: Unifying looking glass querying}.
\newblock In {\em Proc. of PAM\/} (2016).

\bibitem{giotsas2017detecting}
{\sc Giotsas, V., Dietzel, C., Smaragdakis, G., Feldmann, A., Berger, A., and
  Aben, E.}
\newblock {Detecting Peering Infrastructure Outages in the Wild}.
\newblock In {\em Proc. of ACM SIGCOMM\/} (2017).

\bibitem{giotsas2014inferring}
{\sc Giotsas, V., Luckie, M., Huffaker, B., and Claffy, K.~C.}
\newblock {Inferring complex AS relationships}.
\newblock In {\em Proc. of IMC\/} (2014).

\bibitem{giotsas2015mapping}
{\sc Giotsas, V., Smaragdakis, G., Huffaker, B., Luckie, M., and Claffy, K.~C.}
\newblock {Mapping Peering Interconnections to a Facility}.
\newblock In {\em Proc. of ACM CoNEXT\/} (2015).

\bibitem{giotsas2013inferring}
{\sc Giotsas, V., Zhou, S., Luckie, M., et~al.}
\newblock {Inferring Multilateral Peering}.
\newblock In {\em Proc. of ACM CoNEXT\/} (2013).

\bibitem{gupta2014peering}
{\sc Gupta, A., Calder, M., Feamster, N., Chetty, M., Calandro, E., and
  Katz-Bassett, E.}
\newblock {Peering at the Internet's Frontier: A First Look at ISP
  Interconnectivity in Africa}.
\newblock In {\em Proc. of PAM\/} (2014).

\bibitem{holterbach2015quantifying}
{\sc Holterbach, T., Pelsser, C., Bush, R., and Vanbever, L.}
\newblock {Quantifying interference between measurements on the RIPE Atlas
  platform}.
\newblock In {\em Proc. of ACM IMC\/} (2015).

\bibitem{ixp_db}
{\sc IXPDB}.
\newblock {IXP Database}.
\newblock \url{https://www.ixpdb.net/en/ix-f/ixp-database}.
\newblock Accessed: 13.05.2018.

\bibitem{karney2013algorithms}
{\sc Karney, C.~F.}
\newblock Algorithms for geodesics.
\newblock {\em Journal of Geodesy 87}, 1 (2013), 43--55.

\bibitem{katz2006towards}
{\sc Katz-Bassett, E., John, J.~P., Krishnamurthy, A., Wetherall, D., Anderson,
  T., and Chawathe, Y.}
\newblock {Towards IP geolocation using delay and topology measurements}.
\newblock In {\em Proc. of ACM IMC\/} (2006).

\bibitem{keys2013internet}
{\sc Keys, K., Hyun, Y., Luckie, M., and Claffy, K.}
\newblock {Internet-scale IPv4 alias resolution with MIDAR}.
\newblock {\em IEEE/ACM ToN 21}, 2 (2013), 383--399.

\bibitem{li2004first}
{\sc Li, L., Alderson, D., Willinger, W., and Doyle, J.}
\newblock {A first-principles approach to understanding the Internet's
  router-level topology}.
\newblock In {\em ACM SIGCOMM CCR\/} (2004), vol.~34, pp.~3--14.

\bibitem{lodhi2014open}
{\sc Lodhi, A., Dhamdhere, A., and Dovrolis, C.}
\newblock {Open peering by Internet transit providers: Peer preference or peer
  pressure?}
\newblock In {\em Proc. of IEEE INFOCOM\/} (2014).

\bibitem{luckie2014challenges}
{\sc Luckie, M., Dhamdhere, A., Clark, D., Huffaker, B., and Claffy, K.~C.}
\newblock {Challenges in Inferring Internet Interdomain Congestion}.
\newblock In {\em Proc. of ACM IMC\/} (2014).

\bibitem{luckie2016bdrmap}
{\sc Luckie, M., Dhamdhere, A., Huffaker, B., Clark, D., and Claffy, K.~C.}
\newblock {Bdrmap: Inference of borders between IP networks}.
\newblock In {\em Proc. of ACM IMC\/} (2016).

\bibitem{luckie2013relationships}
{\sc Luckie, M., Huffaker, B., Dhamdhere, A., Giotsas, V., and Claffy, K.~C.}
\newblock {AS relationships, customer cones, and validation}.
\newblock In {\em Proc. of ACM IMC\/} (2013).

\bibitem{marder2016map}
{\sc Marder, A., and Smith, J.~M.}
\newblock {MAP-IT: Multipass accurate passive inferences from traceroute}.
\newblock In {\em Proc. of ACM IMC\/} (2016).

\bibitem{mason2001cisco}
{\sc Mason, A.~G., and Newcomb, M.~J.}
\newblock {\em {Cisco secure Internet security solutions}}.
\newblock Cisco press, 2001.

\bibitem{nipper2016}
{\sc Nipper, A.}
\newblock {Remote Peering (with A look at Resellers as well), \nth{29} Euro-IX
  Forum, Krakow, Poland}.
\newblock \url{https://goo.gl/1ynm26}, Nov 2016.
\newblock Accessed: 13.05.2018.

\bibitem{pdb_update_2017}
{\sc Nipper, A.}
\newblock {PeeringDB Update, DENOG 9}.
\newblock
  \url{https://docs.peeringdb.com/presentation/20171123-DENOG9-nipper.pdf}, Nov
  2017.
\newblock Accessed: 13.05.2018.

\bibitem{nomikos2016traixroute}
{\sc Nomikos, G., and Dimitropoulos, X.}
\newblock {traIXroute: Detecting IXPs in Traceroute Paths}.
\newblock In {\em Proc. of PAM\/} (2016).

\bibitem{norton2014}
{\sc Norton, W.~B.}
\newblock {\em The 2014 Internet Peering Playbook: Connecting to the Core of
  the Internet}.
\newblock DrPeering Press, 2014.

\bibitem{DrPeering-remote-peering}
{\sc {Norton, William B}}.
\newblock {The Great Remote Peering Debate}.
\newblock \url{https://goo.gl/yMrELB}, 2012.
\newblock Accessed: 13.05.2018.

\bibitem{Norton-remote-peering}
{\sc {Norton, William B}}.
\newblock {Understanding Remote Peering (presentation)}.
\newblock \url{https://goo.gl/3WruyV}, 2013.
\newblock Accessed: 13.05.2018.

\bibitem{roughan201110}
{\sc Roughan, M., Willinger, W., Maennel, O., Perouli, D., and Bush, R.}
\newblock {10 Lessons from 10 Years of Measuring and Modeling the Internet's
  Autonomous Systems}.
\newblock {\em IEEE JSAC 29}, 9 (2011), 1.

\bibitem{ixpcost}
{\sc Snijders, J., Abdel-Hafez, S., and Strong, M.}
\newblock {IXP megabit per second cost \& comparison}.
\newblock \url{https://goo.gl/3nx1FJ}.
\newblock Accessed: 13.05.2018.

\bibitem{souquet2016}
{\sc Souquet, S.}
\newblock {France-IX reseller programme - Challenges and evolution after 4
  years, \nth{29} Euro-IX Forum, Krakow, Poland}.
\newblock \url{https://goo.gl/jRXs4b}, Nov 2016.
\newblock Accessed: 13.05.2018.

\bibitem{steenbergen2009practical}
{\sc Steenbergen, R.~A.}
\newblock A practical guide to (correctly) troubleshooting with traceroute.
\newblock {\em NANOG\/} (2009), 1--49.

\bibitem{Stocker2016Content}
{\sc Stocker, V., Smaragdakis, G., Lehr, W., and Bauer, S.}
\newblock {Content may be King, but (Peering) Location matters: A Progress
  Report on the Evolution of Content Delivery in the Internet}.
\newblock In {\em Proc. of ITS Europe\/} (2016).

\bibitem{teixeira2004dynamics}
{\sc Teixeira, R., Shaikh, A., Griffin, T., and Rexford, J.}
\newblock {Dynamics of hot-potato routing in IP networks}.
\newblock In {\em ACM SIGMETRICS PER\/} (2004), vol.~32, pp.~307--319.

\bibitem{trammell2018revisiting}
{\sc Trammell, B., and K{\"u}hlewind, M.}
\newblock {Revisiting the Privacy Implications of Two-Way Internet Latency
  Data}.
\newblock In {\em Proc. of PAM\/} (2018).

\end{thebibliography}

\end{document}